\documentclass[a4paper,12pt]{article}

\usepackage{geometry}
\usepackage{booktabs,dcolumn}
\usepackage{chngcntr}
\usepackage{psfrag,graphicx}
\usepackage{eso-pic}
\usepackage{amsmath, bm}
\usepackage{longtable}
\usepackage{pdflscape}
\usepackage{appendix}
\usepackage{dashbox}
\usepackage{natbib}
\usepackage{xcolor,colortbl}
\usepackage{tabularx}
\usepackage{ulem}
\usepackage{float}
\usepackage[section]{placeins}

\usepackage[hidelinks,colorlinks,linkcolor=blue!60!black,citecolor=blue!60!black,urlcolor=blue!60!black]{hyperref}

\newcolumntype{z}[1]{D{.}{.}{#1}}

\newcommand{\cb}{\textcolor{blue}}
\newcommand{\cg}{\cellcolor[HTML]{C0C0C0}}

\newcommand{\caviarintp}{{\omega}_c}
\newcommand{\caviararchp}{{\gamma}_c}
\newcommand{\caviargarchp}{{\beta}_c}

\newcommand{\garchintp}{\omega}
\newcommand{\garcharchp}{\gamma}
\newcommand{\garchgarchp}{\beta}

\newcommand{\var}{VaR} 
\newcommand{\q}{Q}     
\newcommand{\es}{\mu}  

\newcommand{\bs}[1]{\boldsymbol{#1}}

\counterwithout{figure}{section}
\counterwithout{table}{section}

\date{}
\begin{document}

\title{

\begin{center} {\Large \bf Using quantile time series and historical simulation to forecast financial risk multiple steps ahead} \end{center}}



\author{Richard Gerlach$^{1}$\footnote{Corresponding author. Email: storti@unisa.it.}, Antonio Naimoli$^{2}$\footnote{Email: anaimoli@unisa.it.}, Giuseppe Storti$^{2}$\footnote{Email: storti@unisa.it.}  
\\
$^{1}$Discipline of Business Analytics, The University of Sydney\\
$^{2}$ Department of Economics and Statistics, University of Salerno}    

\date{} \maketitle

\begin{abstract}
\noindent
A method for quantile-based, semi-parametric historical simulation estimation of multiple step ahead Value-at-Risk (VaR) and Expected Shortfall (ES) models is developed. It uses the quantile loss function, analogous to how the quasi-likelihood is employed by standard historical simulation methods. The returns data are scaled by the estimated quantile series, then resampling is employed to estimate the forecast distribution one and multiple steps ahead, allowing tail risk forecasting. The proposed method is applicable to any data or model where the relationship between VaR and ES does not change over time and can be extended to allow a measurement equation incorporating realized measures, thus including Realized GARCH and Realized CAViaR type models. Its finite sample properties, and its comparison with existing historical simulation methods, are evaluated via a simulation study. A forecasting study assesses the relative accuracy of the 1\% and 2.5\% VaR and ES one-day-ahead and ten-day-ahead forecasting results for the proposed class of models compared to several competitors.





\vspace{0.5cm}


\vspace{0.5cm}
\end{abstract}

\newpage

{\centering
\section{\normalsize INTRODUCTION}\label{introduction_sec}
\par
}
Since its introduction by J.P. Morgan in 1993, Value-at-Risk (VaR) is widely employed by financial institutions for capital allocation and risk management purposes. Let $\mathcal{I}_t$ be the information available at time $t$ and 
\[
F_{t}(r)=Pr(r_{t}\leq r | \mathcal{I}_{t-1})
\]
be the Cumulative Distribution Function (CDF) of the return $r_{t}$ conditional on $\mathcal{I}_{t-1}$. Assuming that $F_{t}(.)$ is strictly increasing and continuous on the real line ${\rm I\!R}$, VaR is the negative of the one-step-ahead $\alpha$ level quantile at time $t$, defined as:
\begin{equation} \label{var_def}
\q_{\alpha,t}=F^{-1}_{t}(\alpha),\qquad 0 <\alpha <1. \nonumber
\end{equation}
VaR cannot measure typical loss for violations and is not mathematically coherent, i.e., it can favor non-diversification. Expected Shortfall (ES), proposed by \cite{artzner1997} and \cite{artzner1999}, gives the expected loss, conditional on returns exceeding a VaR threshold, and is a coherent risk measure; it is also widely employed for tail risk measurement and is now favored by the Basel Committee on Banking Supervision. Under the same assumptions, the one-step-ahead $\alpha$ level ES can be shown \citep[see][among others]{AceTas2002} to equal the negative of the tail conditional expectation of $r_t$, i.e.:
\[
\text{ES}_{\alpha,t}= -E(r_t|r_t\leq \q_{\alpha;t}, \mathcal{I}_{t-1}).
\]
Historical simulation (HS) for tail risk forecasting is well known. However, standard HS methods suffer from theoretical shortcomings that limit their ability to accurately predict risk under realistic dependence regimes, such as volatility clustering. A popular extension of the HS method uses quasi-maximum likelihood (QML) estimation of GARCH-type volatility models to first filter the return data, while applying the HS method only in a second step to estimate the quantile of the QML standardized residuals, see for example \cite{Gao_Song_2008}. This is referred to as filtered HS (FHS). This paper extends the FHS idea by considering filtering via conditional quantile models, to be denoted as Quantile FHS (QFHS). 

\cite{manganelli2004comparison} develop a method that we consider a pre-cursor to our proposed QFHS methods. They employed standard CAViaR type models \citep{caviar}, estimated by minimising the quantile loss function, to estimate VaR; then regressed the returns exceeding the estimated VaR series on the VaR series itself, to estimate ES \citep[see Equations 9.23-9.25 in][]{manganelli2004comparison}. This method is equivalent to our QFHS approach for $h=1$ period forecasts, but cannot be employed for multi-step ahead forecasting. 

More recently, relying on the theoretical results on the elicitability of the pair (VaR, ES) derived by \cite{Fissler2016}, \cite{tayl2019}, \cite{pattonetal2019},  \cite{mitrodima23} and \cite{chenetal2023} have proposed joint dynamic semi-parametric models for VaR and ES, often called ES-CAViaR models. The last two contributions also allow for the  modelling of long range dependence in VaR and ES dynamics.
However, none of these models, as with the original CAViaR model, can be used to generate multi-step-ahead risk forecasts. 

This paper develops and extends both the QML-based FHS method and the aforementioned method in \cite{manganelli2004comparison} in three different directions. First, we illustrate that regression-based ES estimation is equivalent to a HS method, based on scaling returns by the estimated VaR series. Then, we extend the method so that VaR and ES forecasting can be done at any quantile level, independent and separate to the choice of risk level used for estimating the CAViaR model. Simulations are conducted and presented here, to investigate whether there are more optimal quantile levels for estimation, and how this quantile loss based estimation compares to the standard quasi-likelihood estimation in existing FHS methods, e.g. see \cite{Gao_Song_2008} and \cite{francq2015risk}. 


%
Third, we further extend our proposed QFHS method, employing resampling to allow multi-step ahead forecasting for VaR and ES, via sampling from the full, implied multi-step ahead return distribution. This is an important feature, especially considering the focus of the Basel Accord on 10 day ahead risk forecasting. 

We illustrate how realized measures can be incorporated into the proposed method, by employing the pseudo-likelihood developed by \cite{gerlach_wang_2020} and \cite{peiris2024} in their (log) Realized CAViaR model(s). Finally, we show that the proposed method implicitly assumes, and is applicable to, any data or model where the relationship between VaR and ES in the conditional return distribution does not change over time; which includes the case where the conditional return distribution is constant. Thus, the proposed method can be applied to most modern financial time series models and return datasets. 

In summary, the main contribution of this paper is the proposal of a new class of quantile-based filtered HS methods for estimating the one-step-ahead and multi-step-ahead conditional return distribution, for application to VaR and ES forecasting. The finite sample properties of this method are assessed via a simulation study and compared to existing FHS approaches. In the empirical study, the proposed methods are assessed via their VaR and ES forecasting performance and compared to a range of existing methods, illustrating generally favourable results.

The paper is organized as follows. Section \ref{method_review} briefly provides some background, reviewing existing methods related to our proposed method. Section \ref{proposed_section} proposes and describes the new methods, including the associated quantile loss and quasi-likelihood employed for parameter estimation. Section \ref{simulation} summarizes the simulations conducted. Section \ref{data_empirical_section} presents the forecasting study. Section \ref{conclusion_section} concludes the paper and discusses future work.

\vspace{0.5cm}
{\centering
\section{\normalsize BACKGROUND} \label{method_review}
}
\noindent

HS for tail risk forecasting is well known, though standard methods tend to not accurately predict risk under volatility clustering. The most successful extension of the HS method uses QML estimation of GARCH-type volatility models to first filter the return data, while applying the HS method only in a second step to estimate the quantile of the QML standardized residuals, see for example \cite{Gao_Song_2008}. This is known as filtered HS (FHS). The main steps of the FHS estimation procedure are briefly recalled below.

Consider a 0-mean GARCH(1,1) process:
	\begin{eqnarray}
		r_t &= & \sigma_t \epsilon_t \,\,;\,\, \epsilon_t \sim \text{i.i.d.}\, D(0,1) 
                \label{e:garch1}
\\
		\sigma_t^2 &= &\omega + \gamma r_{t-1}^2 +
		\beta \sigma^2_{t-1}
        \label{e:garch2}
	\end{eqnarray}
where the CDF of $\epsilon_t$ (i.e. of $D(\mu,\sigma)$), is assumed to be strictly increasing and continuous on the real line, with zero mean and unit variance. GARCH FHS (G-FHS) proceeds by estimating the model above via QML, where the likelihood is approximated by a Gaussian distribution for $D$ at the optimization stage. The QML estimates are employed to estimate the iid standardized error series: $\hat{\epsilon}_t = \frac{r_t}{\hat{\sigma}_t}$. Then, the sample $\alpha$-level VaR and ES for the series $\hat{\epsilon}_t$, or a re-sampled version of it, is employed to forecast conditional VaR and ES for the return series, via:
\begin{eqnarray*}
\widehat{\var}_{\alpha,T+1} &=& -\hat{\sigma}_{T+1} \widehat{\q}_{\alpha;\hat{\epsilon}} \\
    \widehat{ES}_{\alpha,T+1} &=&
    -\hat{\sigma}_{T+1}\widehat{\es}_{\alpha;\hat{\epsilon}},
\end{eqnarray*}
where $\widehat{Q}_{\alpha;\hat{\epsilon}}$ is the sample $\alpha$-level quantile for the series $\hat{\epsilon}_t$, and $\widehat{\mu}_{\alpha;\hat{\epsilon}}$ is the sample tail mean for $\hat{\epsilon}_t<\widehat{\q}_{\alpha;\hat{\epsilon}}$,  and $\hat{\sigma}_{T+1}$ is the one-step ahead forecast from the GARCH model. The procedure can be immediately extended to consider alternative non-parametric estimators of ${\q}_{\alpha;\hat{\epsilon}}$ \citep[see][among others]{gourieroux2000}  and $\es_{\alpha;\hat{\epsilon}}$
\citep[see][among others]{chen2008,scaillet2004}. In fact, the volatility model does not need to be a GARCH, it can be any volatility model that has a constant conditional return distribution (i.e. $D$ does not change over time) and can be estimated via QML, including Realized GARCH \citep{hansen2012realized}. Further, re-sampling of the series $\hat{\epsilon}_t$ can also be used to generate multi-step-ahead forecasts for VaR and ES in this FHS framework. 

CAViaR models are proposed by \cite{engle_manganelli_2004} as a general class of dynamic semi-parametric models for predicting VaR. \cite{manganelli2004comparison} extended the CAViaR framework proposing a semi-parametric method for estimating VaR and ES. They employed standard CAViaR-type models, estimated by minimizing the quantile loss function, to estimate VaR and forecast it one-step-ahead. They proposed forecasting ES one-step-ahead via regressing the returns exceeding the estimated quantile series on that quantile series, as in Equations (9.23) and (9.25) in \cite{manganelli2004comparison}, now replicated:
\begin{eqnarray}\label{em2325}
r_t &=& \delta_{\alpha} \q_{\alpha,t} + \eta_t \,\,\,\, \text{for} \,\,\,\, r_t < \q_{\alpha,t} 
\label{e:eq1}\\  
\hat{E}_t (r_t | r_t < \q_{\alpha,t}) &=& \hat{\delta}_{\alpha} \widehat{\q}_{\alpha,t}   \, ,
\label{e:eq2}
\end{eqnarray}
where 
$\widehat{\q}_{\alpha,t}$ is estimated via a CAViaR model and Equation \ref{e:eq2} gives the {(negative)} ES estimator, where ordinary least squares (OLS) is used to estimate $\delta_{\alpha}$. 

This method can only be used to estimate ES at the same level $\alpha_0$ that $\widehat{\var}_{\alpha_0,t}$ was estimated at and only for $h=1$ step forecasting. In Section \ref{proposed_section} an analogy between this method and our proposed QFHS method is shown and then the method is extended in order to generate multi-step forecasts. 

\vspace{0.5cm}
{\centering
\section{\normalsize PROPOSED METHOD: THE QUANTILE FILTERED HISTORICAL SIMULATION (QFHS)}\label{proposed_section}
}


In this section we illustrate our proposed method, which will be referred to as Quantile Filtered Historical Simulation (QFHS), as it extends the FHS approach to a quantile regression framework. First, a time series model for the dynamic $\alpha$-level quantile $\q_{\alpha,t}=-\var_{\alpha,t}$ should be specified, e.g. CAViaR, where we define:
	\begin{eqnarray*}
		\alpha = Pr( r_{t} < \q_{\alpha,t} | \mathcal{F}_{t-1} )  \,\,\,;\,\, \mbox{ES}_{\alpha,t} = -E\left[ r_{t} | r_{t} < \q_{\alpha,t},\mathcal{F}_{t-1} \right]
	\end{eqnarray*} 
	which implies:
	\begin{eqnarray*}
		\alpha = Pr\left( \epsilon_{\alpha,t} < -1 | \mathcal{F}_{t-1} \right)  \,\,\,;\,\, \mbox{ES}_{\alpha,t} = \var_{\alpha,t} E\left[ \epsilon_{\alpha,t} | \epsilon_{\alpha,t} < -1,\mathcal{F}_{t-1} \right]
	\end{eqnarray*} 
	where $\epsilon_{\alpha,t} = \frac{r_{t}}{-\q_{\alpha,t}}$. Under correct specification of the quantiles, the series $\epsilon_{\alpha,1},\ldots,\epsilon_{\alpha,T}$ is an iid series with $\alpha$-level quantile of -1, as also noted by \cite{manganelli2004comparison}. If we further assume $\q_{\alpha,t} = \sigma_{t} c_{\alpha}$, for some constant $c_\alpha$, it follows that $E(\epsilon_{\alpha,t}) = 0$. The QFHS predictor takes resamples of the estimated series $\epsilon_{\alpha,t}$:
 \begin{equation}
\hat{\bm{\epsilon}}_{\alpha,t} = \frac{r_{t}}{-\widehat{\q}_{\alpha,t}},\quad t=1,\ldots, T     
\label{e:qfhs_pred}
 \end{equation}

\noindent as the main ingredient in generating multi-step forecasts of ${\var}_{\alpha,t}$ through the following procedure
 \begin{itemize}
     \item[\textbf{Step 1:}] Calculate 1 step forecast $\widehat{\var}_{\alpha,T+1} | \mathcal{F}_T$ from the estimated quantile model
     \item[\textbf{Step 2:}]  For $j$ in $1:M$: 
    \begin{enumerate} 
			\itemsep 10pt
			\item For $h=1$, sample one unit from $\hat{\bm{\epsilon}}_{\alpha}$, label it $\hat{\epsilon}_{\alpha, T+1}^{[j]}$
			\item Set $r_{T+1}^{[j]} = \widehat{\var}_{\alpha,T+1} \, \hat{\epsilon}_{\alpha, T+1}^{[j]}$ 
   \item For $t$ in $2:h$: 
			\begin{enumerate} 
				\item Given $r_{T+t-1}^{[j]}$ calculate 1 step forecast $ \widehat{\var}_{\alpha,T+t}$
				\item Set $r_{T+t}^{[j]} = \widehat{\var}_{\alpha,T+t} \, \hat{\epsilon}_{\alpha,T+t}^{[j]}$
			\end{enumerate}
		\end{enumerate}
  \item[\textbf{Step 3:}]  Consider $r_{T+1}^{[j]},\ldots,r_{T+h}^{[j]}$, which is a realization from $r_{T+1},\ldots,r_{T+h} | \mathcal{F}_T$ with CAViaR quantile dynamics, and compute cumulated $h$ steps simulated returns as
  \[
R^{[j]}_{\alpha,T+h}=\sum_{j=1}^{h}r_{T+h}^{[j]},\quad j=1,\ldots,M.
  \]
  which is a realization from $R_{\alpha,T+h}|\mathcal{F}_T$
  \item[\textbf{Step 4:}] $h$ step VaR ($\widehat{\var}_{\alpha, T+h|T}$) and ES ($\widehat{ES}_{\alpha, T+h|T}$) at the $\alpha$ level are finally computed as the negative sample quantile and tail expectation of $$R^{[1]}_{\alpha,T+h},\ldots,R^{[M]}_{\alpha,T+h},$$
  respectively.

 \end{itemize}
When taking $h=1$, employing $M=T$ and setting each $\hat{\epsilon}_{\alpha, T+1}^{[j]} = \hat{\epsilon}_{\alpha, j}$, this procedure is equivalent to that for forecasting ES one-step-ahead in \cite{manganelli2004comparison}. This can be seen by dividing Equation \ref{e:eq1} through by $\q_{\alpha, t}$, so that $\hat{\delta}_{\alpha}$ is the sample tail average of $\frac{r_t}{\widehat{\q}_{\alpha, t}}$, which when multiplied by the (constant) quantile forecast becomes equivalent to our $h=1$ QFHS estimator. 

The proposed QFHS procedure can also be applied to forecast conditional volatility, e.g. the forecast conditional standard deviation is simply the standard deviation of the sample $R^{[1]}_{\alpha,T+h},\ldots,R^{[M]}_{\alpha,T+h}$.

In the illustration of the QFHS procedure, we set the \emph{target} VaR level in Step 4 for the computation of sample quantile and tail average, say $\alpha_0$, equal to the quantile level used for the specification and estimation of the underlying CAViaR model, say $\alpha_{\text{est}}$. However, it is immediately seen that the QFHS can be implemented for forecasting $\widehat{\var}_{\alpha_0, T+h|T}$ and $\widehat{ES}_{\alpha_0, T+h|T}$ setting $\alpha_{\text{est}} \neq \alpha_{0}$. The next section attempts to shed light on the selection of $\alpha_{\text{est}}$ \c{via simulations}.

We then focus on a comparative analysis of the FHS and QFHS methods. Although the QFHS and the FHS methods follow a similar logic, there are some notable differences between the two approaches. First, in FHS, the residuals are obtained under the unit variance identification restriction. In the QFHS, this is replaced by an identification constraint imposed on the quantile of the residuals at the $\alpha_{\text{est}}$ level, which by construction is forced to be equal to -1. Since our main purpose in the current framework is to forecast a conditional quantile, this choice seems more natural than the variance restriction typically imposed in GARCH models. Second, as shown by \cite{komunjer2005}, the estimation of CAViaR models has a QML interpretation. 
Thus, from this perspective, the QFHS and FHS could be seen as special cases of a more general procedure where the filtering step is performed by means of a consistent QML estimator obtained by optimising a Gaussian QL in the FHS case and an asymmetric Laplace QL function in the QFHS case. In this respect, the QFHS is more flexible than the FHS, since different choices of $\alpha_{\text{est}}$ lead to different asymmetric Laplace QL functions that can potentially be used to generate consistent risk forecasts. In addition, a further generalisation of the QFHS estimator could be obtained by replacing the asymmetric Laplace QL function corresponding to the standard quantile regression estimator with the more general family of tick exponential QL functions whose properties are derived and discussed by \cite{komunjer2005}.


\vspace{0.5cm}

{\centering
\section{\normalsize SIMULATION AND CHOICE OF QUANTILE LEVEL} \label{simulation}
}
\noindent

In this section, we investigate the empirical performance of the QFHS via Monte Carlo simulations. First, in Section \ref{ss:predsim} we present the results of a simulation study aimed at assessing the performance of the QFHS in risk forecasting for different values of $\alpha_{\text{est}}$ and $\alpha_0$ and different forecasting horizons. Second, Section \ref{ss:secaviar} illustrates an additional simulation study whose aim is to provide some theoretical guidance on the selection of $\alpha_{\text{est}}$. 

\subsection{Predictive accuracy of the QFHS: simulation evidence}
\label{ss:predsim}
As discussed in Section \ref{proposed_section}, for a given target risk level $\alpha_0$, the QFHS can be used to predict risk at level $\alpha_0$ starting from several different estimated CAViaR models of order $\alpha_{\text{est}}$. Based on Monte Carlo simulations, this section attempts to provide guidance on the choice of $\alpha_{\text{est}}$ in practical applications, by considering relative accuracy of forecasts of VaR and ES for varying $\alpha_{\text{est}}$. 

Our simulation experiments consider different DGPs obtained as special cases of the GARCH-type framework
\begin{align*}
r_t&=\sigma_t z_t \qquad \epsilon_t \overset{iid}{\sim}D(0,1;\bs{\psi})\\
\sigma^2_t&=\sigma(\mathcal{I}_{t-1};\bs{\gamma})
\end{align*}
where $D$ is chosen as Normal ($N$), Student-t ($t$) or Skewed-Student-t ($st$) \citep{hansen1994autoregressive};  $\bs{\psi}$ denotes a vector of shape parameters depending on the specific distribution $D$. Namely, in the $t$ case, $\bs{\psi}\equiv \nu$, where $\nu$ is the degrees-of-freedom parameter while, for the $st$ distribution,  $\bs{\psi}\equiv (\nu,\xi)$, where $\xi$ is the skewness parameter. Here, negative (positive) values of $\xi$ will correspond to negatively (positively) skewed distributions, respectively, while $\xi=0$ will indicate symmetry. We consider the following set of values for the degrees of freedom $\nu \in \{ 2.1, 2.5, \,5, \, \infty\}$, while the skewness coefficient $\xi \in \{-0.05,\, \,0\, \}$.
It is worth noting that for $\nu=\infty$ and $\xi=0$ we have the Normal distribution, for $\nu=\infty$ and $\xi \neq 0$ we have the Skewed Normal distribution, while for $\xi=0$ we obtain the usual Student-t. 

For our simulation, we assume that the volatility function $\sigma(.|.)$ is a standard GARCH(1,1) model as in Equation \ref{e:garch2} where the parameters are chosen as
\[ \garchintp = 0.01; \quad \garcharchp=0.10; \quad\garchgarchp=0.89 \, . \]
Data is simulated from the specified GARCH(1,1) model, employing each paired combination of values for $\nu, \xi$, for $N = 500$ replications, each of size $T=3000$ observations. Each dataset has true VaR and ES at nominal risk levels $\alpha_0 = 0.025, 0.01$ given by the DGP and its simulated values. 

A GARCH model is estimated by QML for each dataset and a bootstrap sample of 25000 is applied to generate $h=1$ and $h=10$ step forecasts of 
VaR and ES at nominal risk levels $\alpha_0 = 0.025, 0.01$. 

Further, six CAViaR-IG (IG) models are estimated for each of the 500 datasets for each DGP, via minimizing the quantile loss function, each employing one value of: 
\[
\alpha_{\text{est}} \in \{ 0.01, 0.025, 0.05, 0.1, 0.15, 0.2\} \, .
\]
Our proposed procedure in Section \ref{proposed_section} is then employed, using a bootstrap sample of 25000, to generate $h=1$ and $h=10$ step ahead forecasts of 
VaR and ES at nominal risk levels $\alpha_0 = 0.025, 0.01$.

\begin{figure}[h!]
\centering
\caption{Forecast accuracy for VaR, ES from 500 GARCH Gaussian replicates.
}
\includegraphics[width=0.8\textwidth]{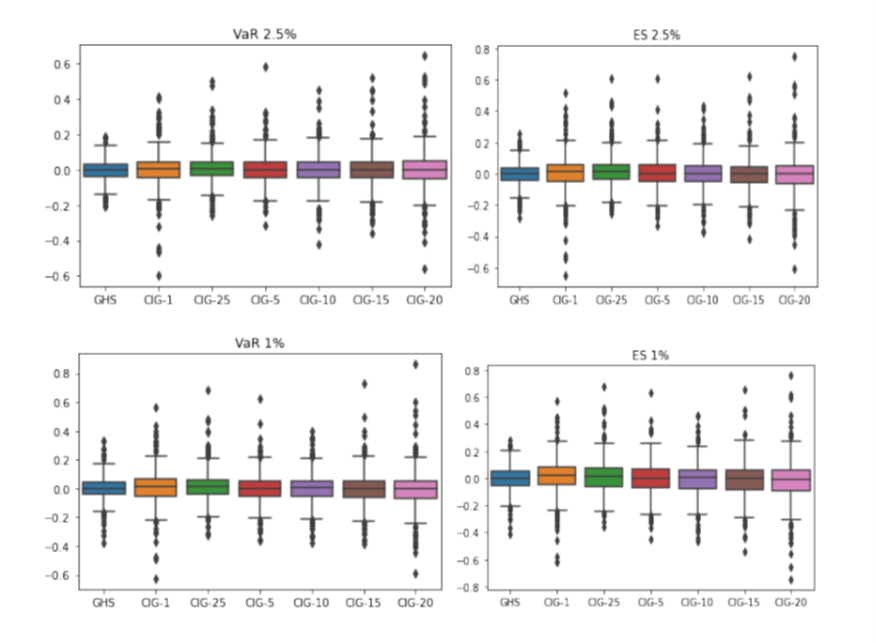}\\
\label{fig:acc_N}

\end{figure}

\begin{figure}[h!]
\centering
\caption{Forecast accuracy for VaR, ES from 500 GARCH-skt($\nu=$ 2.5, $\xi=$ -0.05) replicates.
}
\includegraphics[width=0.8\textwidth]{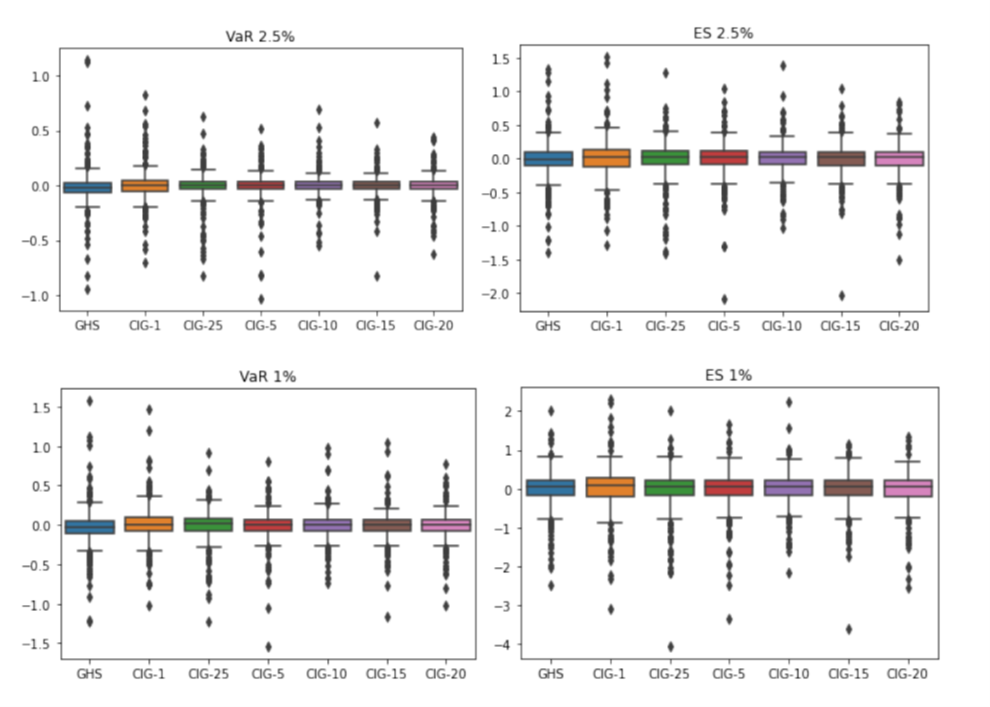}\\
\label{fig:acc_skt25}

\end{figure}

Figures \ref{fig:acc_N} and \ref{fig:acc_skt25} and Table \ref{tab:h1_acc} highlight that QFHS methods don't compare favourably with GARCH-FHS (GHS) when the DGP has Gaussian errors, even when skewness is present, for $h=1$ forecasting. However, as the kurtosis increases, here as $\nu$ decreases, the QFHS methods compare more favourably to, and then outperform, GHS. Further, the most accurate QFHS approach, with typically smallest RMSEs, has $\alpha_{\text{est}} > \alpha_0$, highlighting an advantage of our flexible approach. For $h=1$, generally as kurtosis increases, QFHS with $\alpha_{\text{est}} \in \{0.05, 0.1, 0.15\}$ are the most accurate methods, under both skewed and non-skewed DGPs, whilst FHS becomes less and less relatively accurate. In general FHS outperforms QFHS with $\alpha_{\text{est}} = 0.01$, except under the highest level of kurtosis.

Finally, if one was solely interested in $h=1$ step VaR forecasts at $\alpha_0 = 0.01$ or $0.025$, under the DGPs considered, it is more accurate to estimate the CAViaR model at $\alpha_{\text{est}} = \in \{0.05, 0.1, 0.15\}$ and take the sample $100 \alpha_0$ percentile of a bootstrap sample of the quantile standardised returns (via our QFHS procedure), than it is to estimate the CAViaR model using $\alpha_0 = \alpha_{\text{est}}$ directly. Further, more accurate ES forecasts also result than employing the method in \cite{manganelli2004comparison} that sets $\alpha_0 = \alpha_{\text{est}}$. Hence, even for $h=1$, our method is a useful extension of both \cite{manganelli2004comparison} and the original CAViaR model.

Table \ref{tab:h10_acc} shows the analogous results for $h=10$ step VaR and ES forecasting. The results are similar to Table \ref{tab:h1_acc} in that, again, the most accurate/favoured QFHS approach typically has $\alpha_{\text{est}} > \alpha_0$ and $\alpha_{\text{est}} \in \{0.05, 0.1, 0.15\}$ and that QFHS compares more favourably to FHS as the level of kurtosis increases. Differently to $h=1$, here QFHS with $\alpha_{\text{est}} \in \{0.05, 0.1, 0.15\}$ also outperfoms FHS when only skewness is present. Thus, either skewness or increasing kurtosis in the DGP leads to QFHS with $\alpha_{\text{est}} \in \{0.05, 0.1, 0.15\}$ being the most accurate method, whilst FHS outperforms QFHS with $\alpha_{\text{est}} = 0.01$ in most DGP cases.

\begin{table}[!h]
\setlength{\tabcolsep}{2pt}
\renewcommand{\arraystretch}{0.65}
\small
\caption{ \small Average ranks for $h=1$ step-ahead forecast accuracy via RMSE for $T=3000$; averaged over the four tail risk quantities. Boxes indicate the best average rank, blue indicates 2nd best method and bold indicates the worst average rank.\\}
\centering
\label{tab:h1_acc}
\begin{tabular}{lcccccccc}
\hline
DGP dist. & N & t(5) & t(2.5) & t(2.1) & sk-N & skt(5) & skt(2.5) & skt(2.1) \\
Method & & & & & & & \\ \hline
GHS &\fbox{1.0}&\fbox{1.5}& 5.75 &{\bf 7.0}&\fbox{1.0}&\cb{1.75}& 6.0 &{\bf 7.0} \\ \hline
IG ($\alpha_{\text{est}}$ shown)  & & & & & & & \\
0.01 & 5.5 &{\bf 7.0}&{\bf 7.0}& 5.0 & 4.25 &{\bf 7.0}&{\bf 7.0}& 5.75 \\
0.025 &\cb{2.5}& 6.0 & 5.25 & 6.0 & 3.0 & 6.0 & 5.25 & 4.0   \\
0.05 &\cb{2.5}& 5.0 &\fbox{1.0}&\fbox{1.0}&\cb{2.0}& 4.25 & 4.25 &\fbox{2.0} \\
0.1 & 4.75 & 2.75 & 3.0 & 3.0 & 4.75 &\fbox{1.25}&\fbox{1.5} &\cb{2.5} \\
0.15 & 4.75 &\cb{1.75}& 4.0 &\cb{2.25}& 6.0 & 3.25 &\cb{2.0} &\fbox{2.0} \\ 
0.2 &{\bf 7.0}& 4.0 &\cb{2.0}& 3.75 &{\bf 7.0}& 4.5 & 2.5 & 4.75  \\
\hline
\end{tabular}
\end{table}

\begin{table}[!h]
\setlength{\tabcolsep}{2pt}
\renewcommand{\arraystretch}{0.65}
\small
\caption{ \small Average ranks for $h=10$ step forecast accuracy via RMSE for $T=3000$; averaged over the four tail risk quantities. Boxes indicate the best average rank, blue indicates 2nd best method and bold indicates the worst average rank.\\}
\centering
\label{tab:h10_acc}
\begin{tabular}{lcccccccc}
\hline
DGP dist. & N & t(5) & t(2.5) & t(2.1) & Sk-N & skt(5) & skt(2.5) & skt(2.1) \\
Est. & & & & & & & \\ \hline
GHS &\fbox{1.0}&\fbox{1.25}& 3.25 &{\bf 6.0}&{\bf 6.5}& 3.0 & 4.75 & 4.5 \\ \hline
IG ($\alpha_{\text{est}}$ shown)  & & & & & & & \\
0.01 & 6.0 &{\bf 7.0}&{\bf 7.0}& 5.75 & 5.25 &{\bf 7.0}& 5.25& 6.0  \\
0.025 & 3.0 & 5.5 &\cb{2.75}& 3.0 & 3.0 & 4.75 & 4.5 &{\bf 7.0}    \\
0.05  &\cb{2.0}&\cb{1.75}& 3.0 &\fbox{1.75}&\cb{2.0}& 3.75 &\cb{2.75}& 2.25 \\
0.1 & 4.0 & 3.0 &\fbox{2.5}&\cb{2.0}&\fbox{1.0}&\fbox{1.0}&\fbox{1.25}&\fbox{1.75} \\
0.15  & 5.0 & 4.0 & 4.0 & 5.0 & 4.0 &\cb{2.5}& 3.5&\cb{2.0} \\ 
0.2  &{\bf 7.0}& 5.5 & 5.5 & 4.5 & 6.25 & 6.0 &{\bf 6.0}& 4.5  \\
\hline
\end{tabular}
\end{table}

The next section provides further insight into how the choice of $\alpha_{\text{est}}$ affects QFHS tail risk forecasting. 

\subsection{Statistical insight on the choice of $\alpha_{\text{est}}$: Monte Carlo evidence}
\label{ss:secaviar}


Based on Monte Carlo simulations, this section attempts to provide further guidance on the choice of $\alpha_{\text{est}}$ in practical applications, for a given $\alpha_0$.

In particular, we investigate the dependence of the accuracy of the estimated CAViaR coefficients on the quantile order $\alpha_{\text{est}}$ used for the estimation. The underlying idea is that the accuracy of the risk forecasts, obtained via the QFHS, is likely to be affected by the efficiency of the estimators of CAViaR parameters. In particular, as confirmed by our empirical results on both simulated and real data, we expect better risk forecasts in the QFHS to be driven by more efficient estimators of the CAViaR coefficients, after accounting for the different $\alpha_{\text{est}}$ level.

To start, we remind that the 1-step ahead QFHS VaR predictor is computed as the empirical quantile of the bootstrap sample
\begin{equation}
r^{*}_{T+1}=-\widehat{\q}_{T+1,\alpha}\epsilon^{*}_{\alpha,T+1}
\label{e:qfhs_ret_1}
\end{equation}
where $\alpha$ is the quantile \emph{estimation} level ($\alpha_{\text{est}}$) of the underlying CAViaR model (not necessarily coincident with the \emph{target} VaR level $\alpha_0$), $\epsilon^{*}_{\alpha,T+1}=-r_{t_0}/\widehat{\q}_{t_0,\alpha}$, with $1\leq t_0 \leq T$, for some randomly chosen $t_0$. 
Reminding that Equation \ref{e:garch1} implies that $\q_{t,\alpha}=\sigma_t Q_{\alpha;\epsilon}$, it is easy to show that Equation \ref{e:qfhs_ret_1} can be rewritten as
\begin{equation}
r^{*}_{T+1}={\q}_{\alpha;\epsilon}\widehat{\sigma}_{\alpha,T+1}
\frac{r_{t_0}}{{\q}_{\alpha;\epsilon} \widehat{\sigma}_{\alpha,t_0}}=r_{t_0}\frac{\widehat{\sigma}_{\alpha,T+1}}{ \widehat{\sigma}_{\alpha,t_0}}
\label{e:qfhs_ret_2}
\end{equation}
where $\widehat{\sigma}_{\alpha,t}=\widehat{\q}_{t,\alpha}/{\q}_{\alpha;\epsilon}$ can be interpreted as the semi-parametric volatility estimate implied by the fitted CAViaR model, for known ${\q}_{\alpha;\epsilon}$.
Equation \ref{e:qfhs_ret_2} makes clear that the performance of the QFHS estimator is directly affected by the estimation accuracy of the parameters driving the volatility dynamics. The remainder of this section will attempt to shed some light on the nature of this relationship in the framework of CAViaR models.


\cite{engle_manganelli_2004} investigate the asymptotic properties of the estimators of CAViaR parameters proving their consistency, asymptotic normality, and providing a closed-form expression for their asymptotic variance. Under standard assumptions (referenced as AN1-AN4 in their paper), they show that 
\begin{equation}
\sqrt{T}A^{-1/2}_TD_T(\hat{\boldsymbol{\theta}}-\boldsymbol{\theta}_0) \stackrel{d} {\rightarrow}
N(\mathbf{0},I)\qquad \textrm{as } T\to \infty
\end{equation}
\begin{align*}
A_T&=E \left[ T^{-1}\alpha(1-\alpha)\nabla^{'}f_t({\boldsymbol{\theta}_0})
\nabla f_t({\boldsymbol{\theta}}_0)\right]\\
D_T&=E \left[ T^{-1}\sum_{t=1}^{T}h_t(0|\mathcal{I}_{t-1}) \nabla^{'}f_t({\boldsymbol{\theta}}_0)
\nabla f_t({\boldsymbol{\theta}}_0)\right]
\end{align*}
where $\nabla f_t(\boldsymbol{\theta})$ denotes the ($1\times n_p$) gradient of the quantile function $f_t(\boldsymbol{\theta})=Q_{\alpha,t}(\boldsymbol{\theta})$, with $n_p$ being the number of elements in $\boldsymbol{\theta}$, $\varepsilon_{\alpha,t}=r_t-{\q}_{\alpha,t}(\boldsymbol{\theta}_0)$, $h_t(0|\mathcal{I}_{t-1})$ denotes the conditional density of $\varepsilon_{\alpha,t}$ evaluated at 0. Consistent estimates of $A_T$ and $D_T$ can be obtained by replacing expectations with sample averages, $\boldsymbol{\theta}_0$ with $\hat{\boldsymbol{\theta}}$ and $h_t(0|\mathcal{I}_{t-1})$ with a non-parametric density estimator. The estimated $A_T$ and $D_T$ can then be used to compute estimates of the asymptotic standard errors. In this section, our interest is not in estimating asymptotic standard errors, but in using Monte Carlo simulations to investigate the relationship between the theoretical asymptotic standard errors of the components of $\hat{\boldsymbol{\theta}} $ and the level $\alpha_{\text{est}}$ of the estimated quantile model. Since we are assuming complete knowledge of the data generating process (DGP), the quantity $h_t(0|\mathcal{I}_{t-1})$  can be evaluated at the ``true'' conditional density of $\varepsilon_{t,\alpha}$ implied by the chosen DGP. In particular, our simulation experiments consider different DGPs obtained as special cases of the GARCH-type framework, as described in Section \ref{ss:predsim}.
Namely, for this simulation study, we consider the following set of values for the degrees of freedom $\nu$
\[
\{2.1,\, 2.5, \,5,\,8,\, 10, \, 15,\, 20,\, \infty\}
\] 
while the skewness coefficient $\xi$ is assumed to take values in the range 
\[\{-0.15,\, -0.10, \,-0.05,\, \,0,\, 0.05, \, 0.10,\, 0.15\}.\]
We assume the volatility function to be given by the same standard GARCH(1,1) model considered in section \ref{ss:predsim}.
The coefficients of the equivalent CAViaR-IG specification are easily derived from those of the volatility function as follows
\[
\caviarintp(\alpha)=\garchintp\, m_\alpha\, ; \quad \caviararchp(\alpha)=\garcharchp\,m_\alpha \, ; \quad \caviargarchp=\garchgarchp \, .
\]
where $m_\alpha=Q^2_{\alpha;\epsilon}$.
The sample size for all simulations is $T = 3000$, where the data are simulated using $N=500$ replications from each of the above considered DGPs. For each simulated series,  we compute the asymptotic standard errors of the estimators of the implicit CAViaR parameters for different quantile levels. Namely we consider equally spaced $\alpha$ values for $\alpha\in [0.0025,0.30]$, assuming a step size equal to 0.0025. Let ($\hat{\caviarintp}$, $\hat{\caviararchp}$) be the semi-parametric estimates of the CAViaR coefficients (${\caviarintp}$, ${\caviararchp}$). In order to make the standard errors of the estimated $\caviarintp$ and $\caviararchp$ comparable across different quantile orders, we divide them by the factor $m_\alpha$. Namely, the \emph{relative} standard errors for $\hat{\caviarintp}$ and $\hat{\caviararchp}$ are given by $rse(\hat{\caviarintp}(\alpha))=se(\hat{\caviarintp}(\alpha))/m_\alpha$ and $
rse(\hat{\caviararchp}(\alpha))=se(\hat{\caviararchp}(\alpha))/m_\alpha$, respectively. This is equivalent to compute the standard errors of the parameters of the semi-parametric volatility implied by the fitted CAViaR model. 



\begin{figure}[h!]
\centering
\caption{Relative standard errors for $\hat{\caviarintp}(\alpha)$, $\hat{\caviararchp}(\alpha)$ and $\hat{\caviargarchp}(\alpha)$ with respect to different levels of $\alpha$. The curves are generated for different values of $\nu$ and setting $\xi=-0.05$. Note: values in the plots have been normalized to the range $[0,1]$.\\
}
\includegraphics[width=0.9\textwidth]{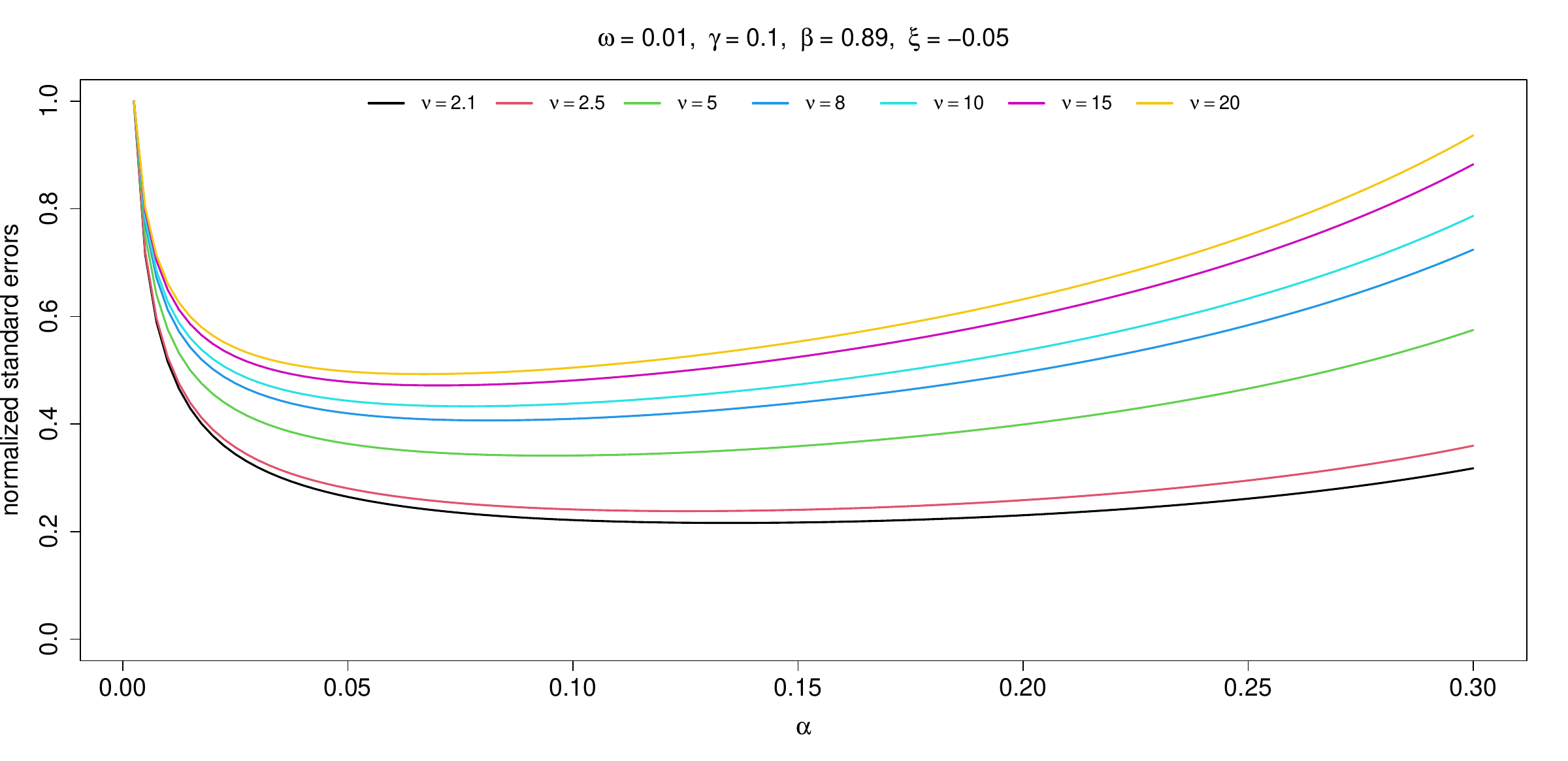}\\
\label{fig:se_norm_omega}

\end{figure}

\begin{figure}[h!]
\centering
\caption{Optimal $\alpha$ value as a function of $\nu$ for fixed $\xi$ (top) and as a function of $\xi$  for fixed $\nu$ (bottom).\\}
\includegraphics[width=\textwidth]{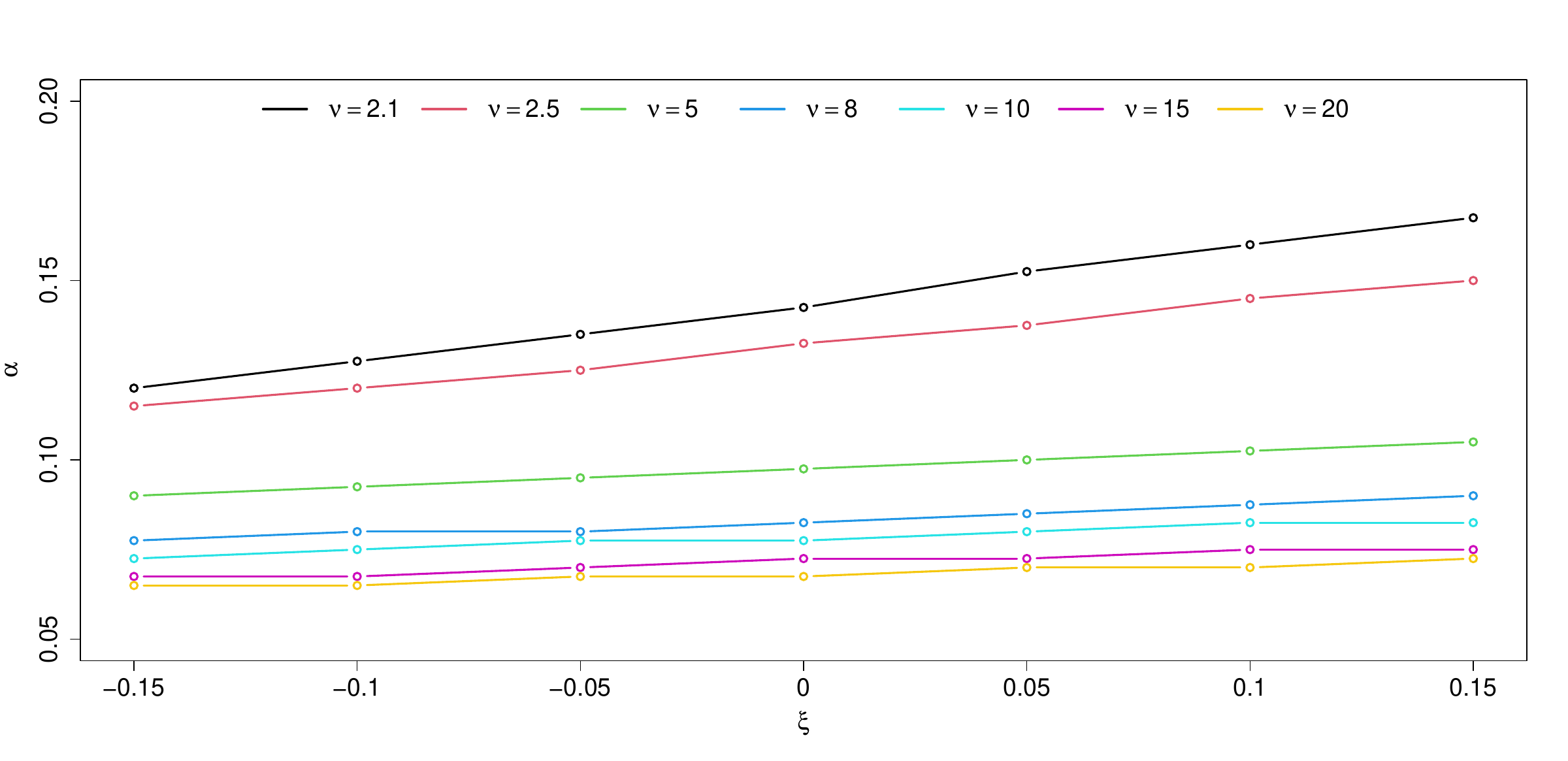}\\
\centering
\includegraphics[width=\textwidth]{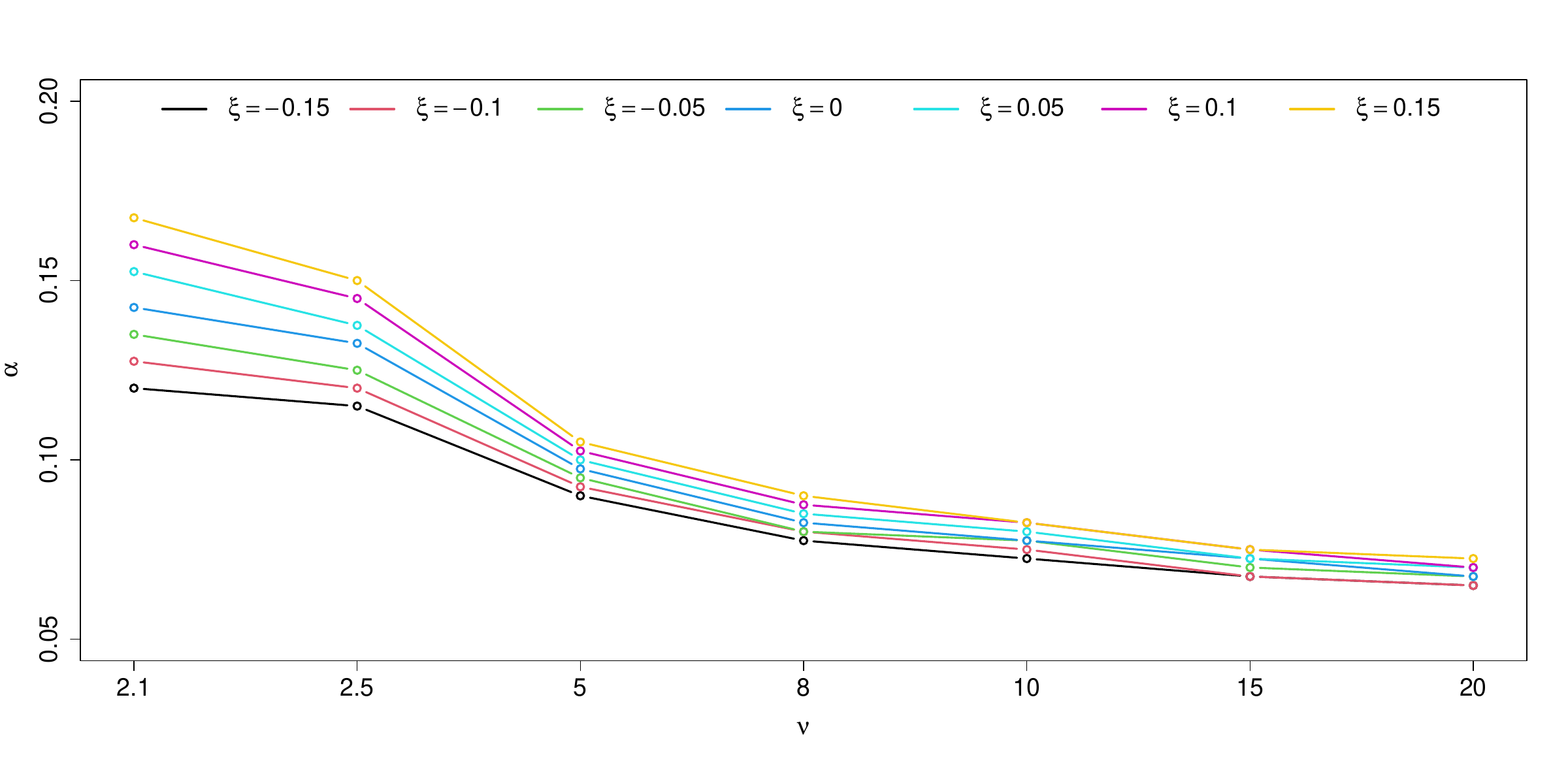}
\label{fig:plot2}
\end{figure}

First, in Figure \ref{fig:se_norm_omega} we investigate the dependence, under different distributional assumptions, of the \emph{rse} of estimated coefficients on the value of $\alpha$. It is worth noting that the asymptotic formulas in \cite{engle_manganelli_2004} yield, for each of the estimated CAViaR coefficients, \emph{rse} patterns that have the same shape and differ only in scale. Normalization of these patterns by their maximum value therefore results in a unique curve rescaled over the interval [0,1]. The plot clearly shows that the \emph{rse} value is dependent on $\alpha$, with the minimum being attained at the same point for all coefficients. Overall, substantial efficiency losses are associated with very low or high values of $\alpha$. Table \ref{t:minalpha_igcaviar} allows to gain deeper insight on the nature of the dependence pattern by reporting the value of $\alpha$ that minimizes the \emph{rse} for the estimated CAViaR parameters under different simulations settings. It is evident that the optimal value of $\alpha$ depends on the distribution, as further confirmed by Figure \ref{fig:plot2} (top panel). For heavy-tailed distributions, the rse is minimized for $\alpha$ values in the range $[0.0900,0.1675]$, for all model parameters. For $\nu \geq 8$, the optimal $\alpha$ takes values in the range $[0.0575,0.0900]$. Overall, these results are largely consistent with the results of the simulation study presented in Section \ref{ss:predsim}.

\begin{table}[h!]
\caption{Level of $\alpha$ that minimizes the relative standard errors (rse) for $\caviarintp(\alpha_i)$, $\caviararchp(\alpha_i)$ and $\caviargarchp(\alpha_i)$ under different combinations of the degrees of freedom $\nu$ and skewness $\xi$. \\}
\centering
\label{tab:param_alpha_min}
\begin{tabular}{c|ccccccc|}
\cline{2-8}
                               & \multicolumn{7}{c|}{$\xi$}                                   \\ \hline
\multicolumn{1}{|c|}{$\nu$}    & -0.15  & -0.10  & -0.05  & 0      & 0.05   & 0.10   & 0.15   \\ \hline
\multicolumn{1}{|c|}{2.1}      & 0.1200 & 0.1275 & 0.1350 & 0.1425 & 0.1525 & 0.1600 & 0.1675 \\
\multicolumn{1}{|c|}{2.5}      & 0.1150 & 0.1200 & 0.1250 & 0.1325 & 0.1375 & 0.1450 & 0.1500 \\
\multicolumn{1}{|c|}{5}        & 0.0900 & 0.0925 & 0.0950 & 0.0975 & 0.1000 & 0.1025 & 0.1050 \\
\multicolumn{1}{|c|}{8}        & 0.0775 & 0.0800 & 0.0800 & 0.0825 & 0.0850 & 0.0875 & 0.0900 \\
\multicolumn{1}{|c|}{10}       & 0.0725 & 0.0750 & 0.0775 & 0.0775 & 0.0800 & 0.0825 & 0.0825 \\
\multicolumn{1}{|c|}{15}       & 0.0675 & 0.0675 & 0.0700 & 0.0725 & 0.0725 & 0.0750 & 0.0750 \\
\multicolumn{1}{|c|}{20}       & 0.0650 & 0.0650 & 0.0675 & 0.0675 & 0.0700 & 0.0700 & 0.0725 \\
\multicolumn{1}{|c|}{$\infty$} & -      & -      & -      & 0.0575 & -      & -      & -      \\ \hline
\end{tabular}%
\label{t:minalpha_igcaviar}
\end{table}



Finally, the bottom panel of Figure \ref{fig:plot2} focuses on the relationship between the optimal $\alpha$ value and the skewness coefficient $\xi$: the effect of skewness is more remarkable for heavy-tailed distributions, while it tends to shrink as the degrees of freedom increase.

In the above experiment, we have assumed knowledge of the error quantile $Q_{\alpha;\epsilon}$. In order to test the robustness of our results in more realistic settings where $Q_{\alpha;\epsilon}$ is estimated, we have performed an additional experiment where, for each $\alpha$, the fitted model is a reparametrized CAViaR-IG
\begin{eqnarray}
\q_{t;\alpha} = Q_{\alpha;\epsilon} \sqrt {\omega+ \gamma r_{t-1}^2 + \beta \sigma_{t-1;\alpha}^2 }
\label{e:repcaviar}
\end{eqnarray}
The parameters of the model in \ref{e:repcaviar} are estimated semi-parametrically by minimizing the quantile loss with respect to ($Q_{\alpha;\epsilon}$,$\gamma$,$\beta$). To ensure identifiability of all the parameters, direct estimation of the intercept $\omega$ is avoided by adopting a variance targeting argument and setting
\[
\omega=(1-\gamma-\beta)s^2_r
\]
where $s^2_r$ denotes the sample variance of returns. Unlike the usual CAViaR models, this alternative parameterization has the advantage of allowing direct semi-parametric estimation of the volatility coefficients $\gamma$ and $\beta$. The Monte Carlo standard deviations of the estimated parameters over $N=1000$ simulated series are used to estimate the standard errors. Thus, compared to the experiment based on the asymptotic results in the first part of this section, this additional analysis makes a twofold contribution. First, it is based on direct estimates of the volatility coefficients, without the need for ad hoc re-scaling. Second, it is based on a finite sample measure of estimation uncertainty rather than asymptotic limit results.

Overall, the new results confirm the general trend that emerges from the analysis of results based on asymptotic standard errors. In order not to detract from the readability of the main text of the paper, the empirical results of this additional experiment have been moved to the Appendix.


\vspace{0.5cm}
{\centering \section{\normalsize EMPIRICAL STUDY} \label{data_empirical_section} }
\noindent
This section presents an application to real stock market data. We consider daily data on the open, high, low and adjusted close of three stock market indices, SP500, FTSE100, ASX200 and three stocks listed on the New York Stock Exchange, Microsoft (MSFT), Gen Digital (GEN) and Agilent Technologies (A). Data is observed from 01/2000 to 09/2023. For each asset close-close log-returns and intra-day (log) ranges are calculated. For $h = 1$, the in-sample period is chosen as 07/2006 - 10/2017 while the forecast period is 10/2017-09/2023. 
This gives an in-sample size of $T \approx 3000$ and $K \approx 1500$ daily returns to forecast. For $h = 10$, in-sample period is 01/2000 - 12/2008,
while the forecast period is 12/2008-09/2023. This gives $T \approx 2270$ and $K \approx 370$ non-overlapping 10-day periods to forecast. The aim of our empirical analysis is twofold. First, in Section \ref{ss:boot}, we extend to a real data framework the analysis performed in \ref{ss:secaviar} in order to investigate the relationship linking the estimation accuracy of volatility dynamic coefficients to the value of $\alpha_{\text{est}}$. To this purpose, we apply a residual Bootstrap resampling algorithm to a re-parameterization of the baseline CAViaR model incorporating variance targeting. Then, in Section \ref{ss:riskfor}, we present the results of an out-of-sample risk forecasting comparison.

\vspace{0.5cm}
{\centering \subsection{\normalsize Bootstrap identification of the optimal $\alpha$} \label{ss:boot} }
\noindent
Under the framework in Section \ref{method_review}, we can write
\[
r_t=\sigma_{t}\epsilon_{t}=-\var_{\alpha;t}\epsilon_{\alpha;t}
\]
where the $\epsilon_{\alpha;t}=-\epsilon_t/\var_{\alpha,t}$ are iid and have a conditional VaR equal to -1. Without loss of generality, for the sake of illustration, it is then convenient to assume that $VaR_{t}$ follows an Indirect GARCH(1,1) model
\[
\var_{\alpha;t}=-\sqrt{{\caviarintp}+{\caviararchp}r^2_{t-1}+\caviargarchp \q^2_{t-1,\alpha}}.
\]
In order to separately estimate the coefficients of the volatility dynamics, the Indirect GARCH(1,1) can be alternatively written as in Equation \ref{e:repcaviar}, 
where, as in Section \ref{simulation}, to guarantee that $Q_{\alpha;\epsilon}$ and $(\omega,\gamma)$ are identified,  variance targeting can be used so that the intercept $\omega$ is replaced by $(1-\gamma-\beta)var(r_t)\approx (1-\gamma-\beta)s^2_r$. 
The volatility coefficients of the re-parameterized model can be semi-parametrically estimated for different quantile orders $\alpha_i$ ($i=1,\ldots,m$) by minimizing the relevant quantile loss function, leading to $\hat{\boldsymbol{\theta}}_{\alpha_i}=(\hat{\gamma}_{\alpha_i},\hat{\beta}_{\alpha_i})'$, where $\hat{\gamma}_{\alpha_i}=\hat{\gamma}_c({\alpha_i})$ and $\hat{\beta}_{\alpha_i}=\hat{\beta}_c({\alpha_i})$. The QFHS procedure described in Section \ref{proposed_section} can then be used to compute a set of $B$ bootstrap resamples of the original returns series $(r_1,\ldots,r_T)$. By re-estimating the model for each resampled series, we finally obtain a bootstrap sample from the distribution of $(\hat{\gamma}_{\alpha_i},\hat{\beta}_{\alpha_i})'$:
\[
(\hat{\gamma}_{\alpha_i;b},\hat{\beta}_{\alpha_i;b})', \qquad \textrm{for $i=1,...,m$ and $b=1,\ldots,B$}.
\]
Bootstrap estimates of the standard errors of the elements of $\hat{\boldsymbol{\theta}}_{\alpha_i}$ can then be computed as
\[
\widehat{se}(\hat{\theta}_{\alpha_i;j})=\sqrt{\frac{1}{B}\sum_{b=1}^{B}
(\hat{\theta}_{\alpha_i,b;j}-\hat{\bar{\theta}}_{i;j}))^2}
\]
where $\hat{\theta}_{\alpha_i;j}$ is the $j$-th element of $\hat{\boldsymbol{\theta}}_{\alpha_i}$, $\hat{\theta}_{\alpha_i,b;j}$ is the $j$-th element of its bootstrap replicate $\hat{\boldsymbol{\theta}}_{\alpha_i,b}$, and $\hat{\bar{\theta}}_{\alpha_i;j}=B^{-1}\sum_{b=1}^{B}\hat{\theta}_{\alpha_i,b;j}$ is the bootstrap mean. 

\begin{figure}[tp!]
\centering
\caption{From top to bottom: standard errors of $\beta$ and $\gamma$ at  different levels of $\alpha_{\text{est}}$, for $ASX\,200$, $FTSE\,100$ and $S\&P\,500$. The patterns are generated by using 500 bootstrap replications. Black line: raw data. Red line: loess-smoothed data.}
\includegraphics[width=0.9\textwidth]{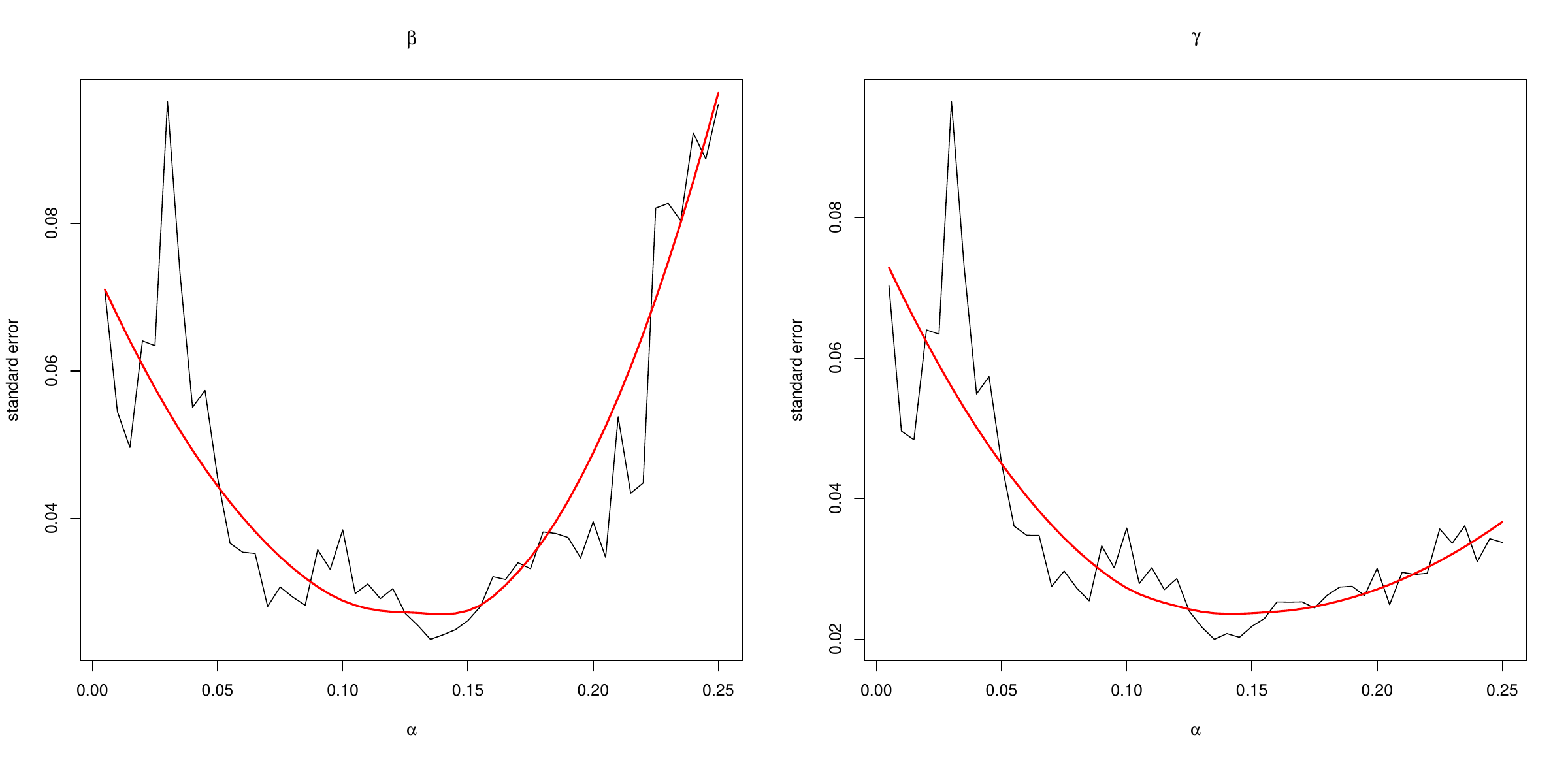}\\
\includegraphics[width=0.9\textwidth]{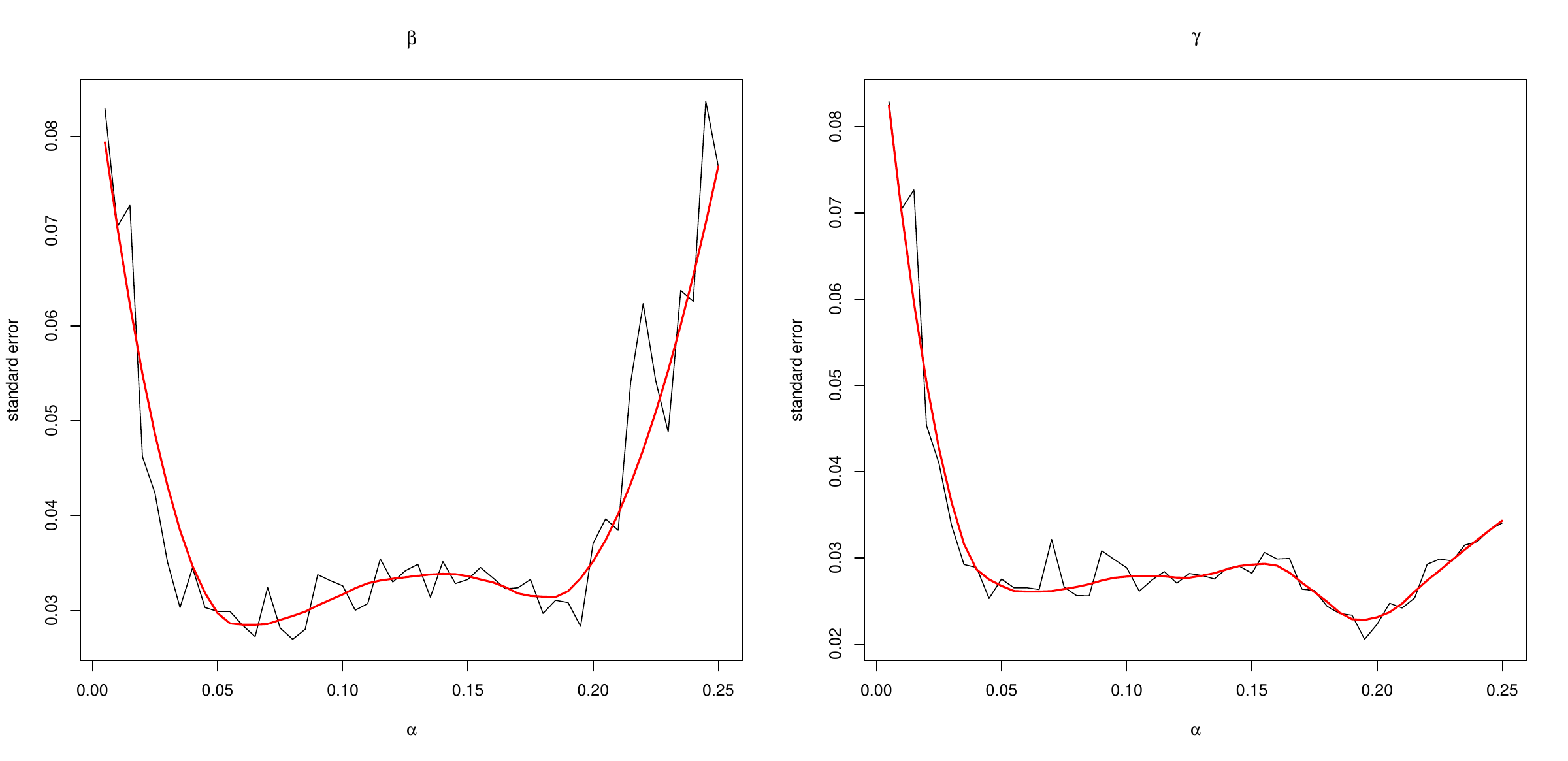}\\
\includegraphics[width=0.9\textwidth]{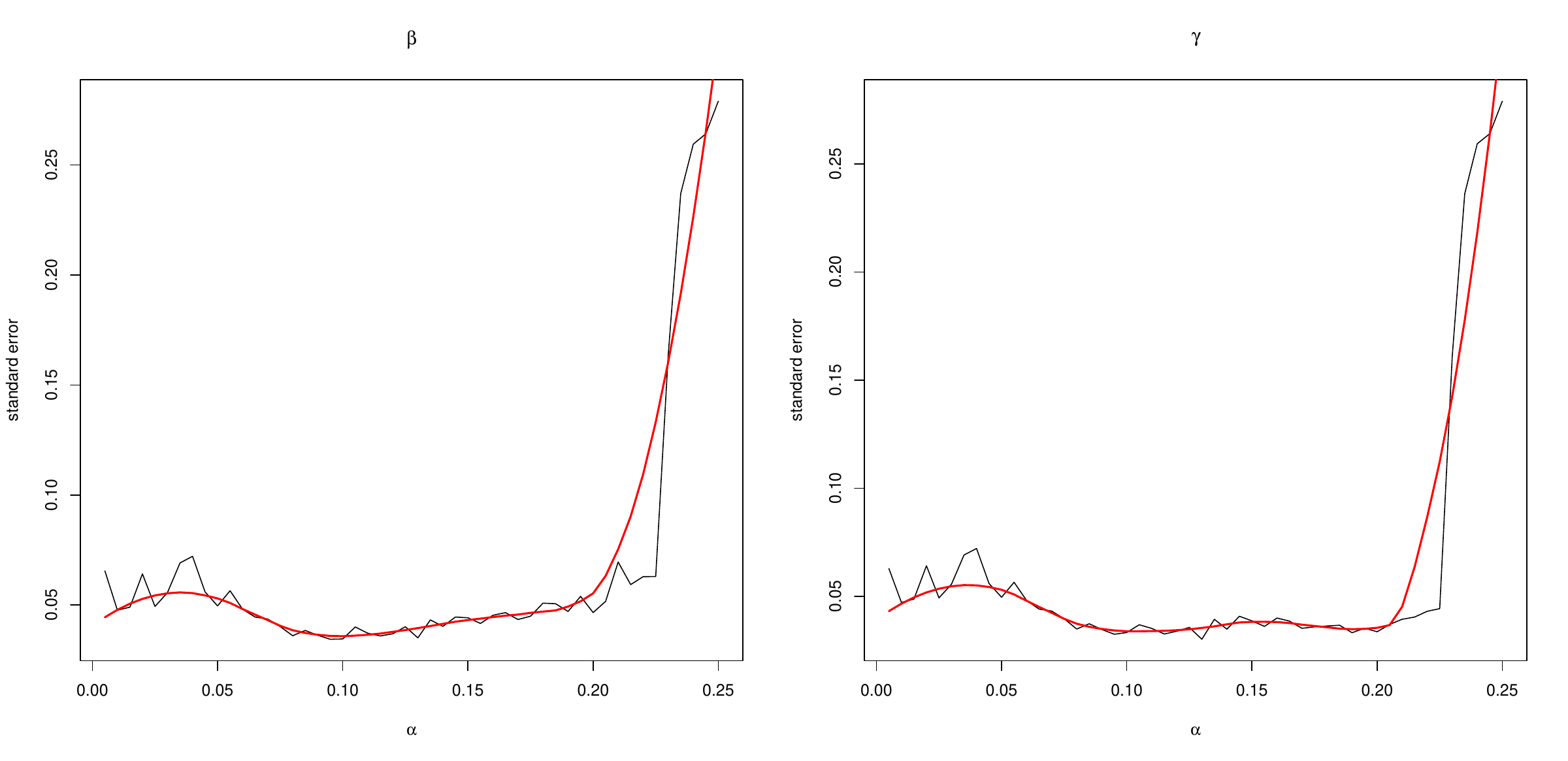}
\label{fig:index_boot}
\end{figure}

\begin{figure}[tp!]
\centering
\caption{From top to bottom: standard errors of $\beta$ and $\gamma$ at  different levels of $\alpha_{\text{est}}$, for $A$, $GEN$ and $MSFT$. The patterns are generated by using 500 bootstrap replications. Black line: raw data. Red line: loess-smoothed data.}
\includegraphics[width=0.9\textwidth]{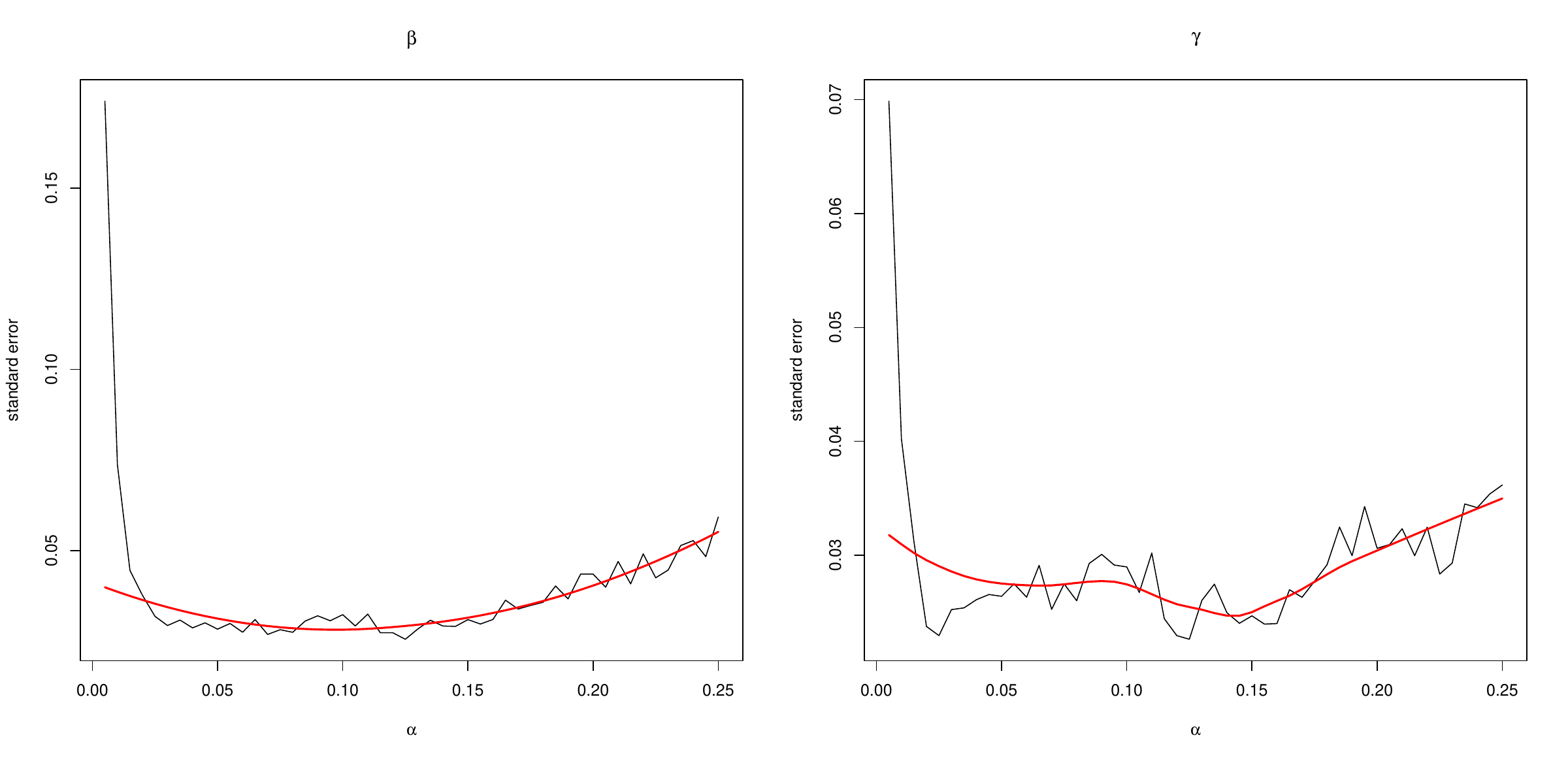}\\
\includegraphics[width=0.9\textwidth]{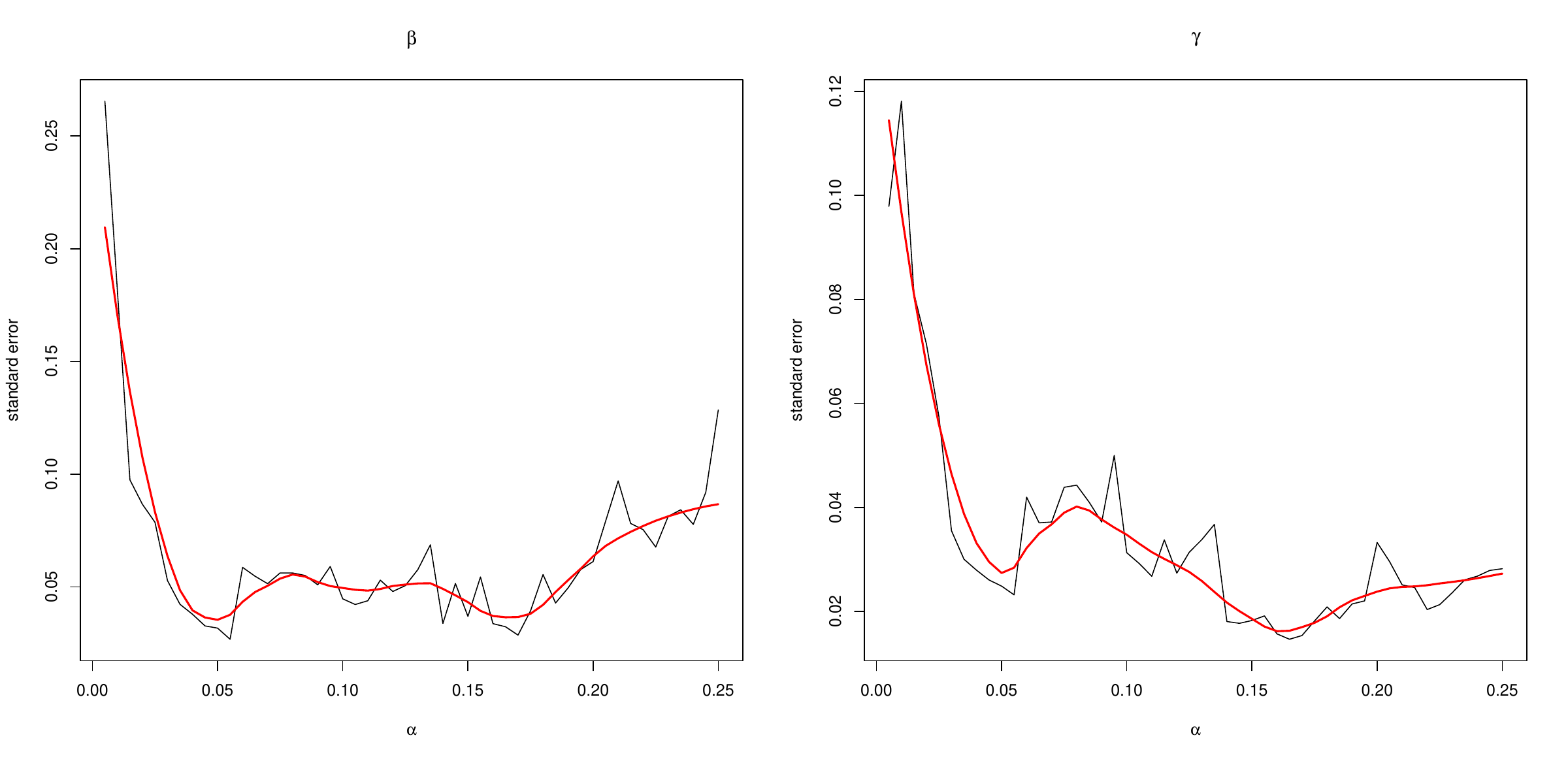}\\
\includegraphics[width=0.9\textwidth]{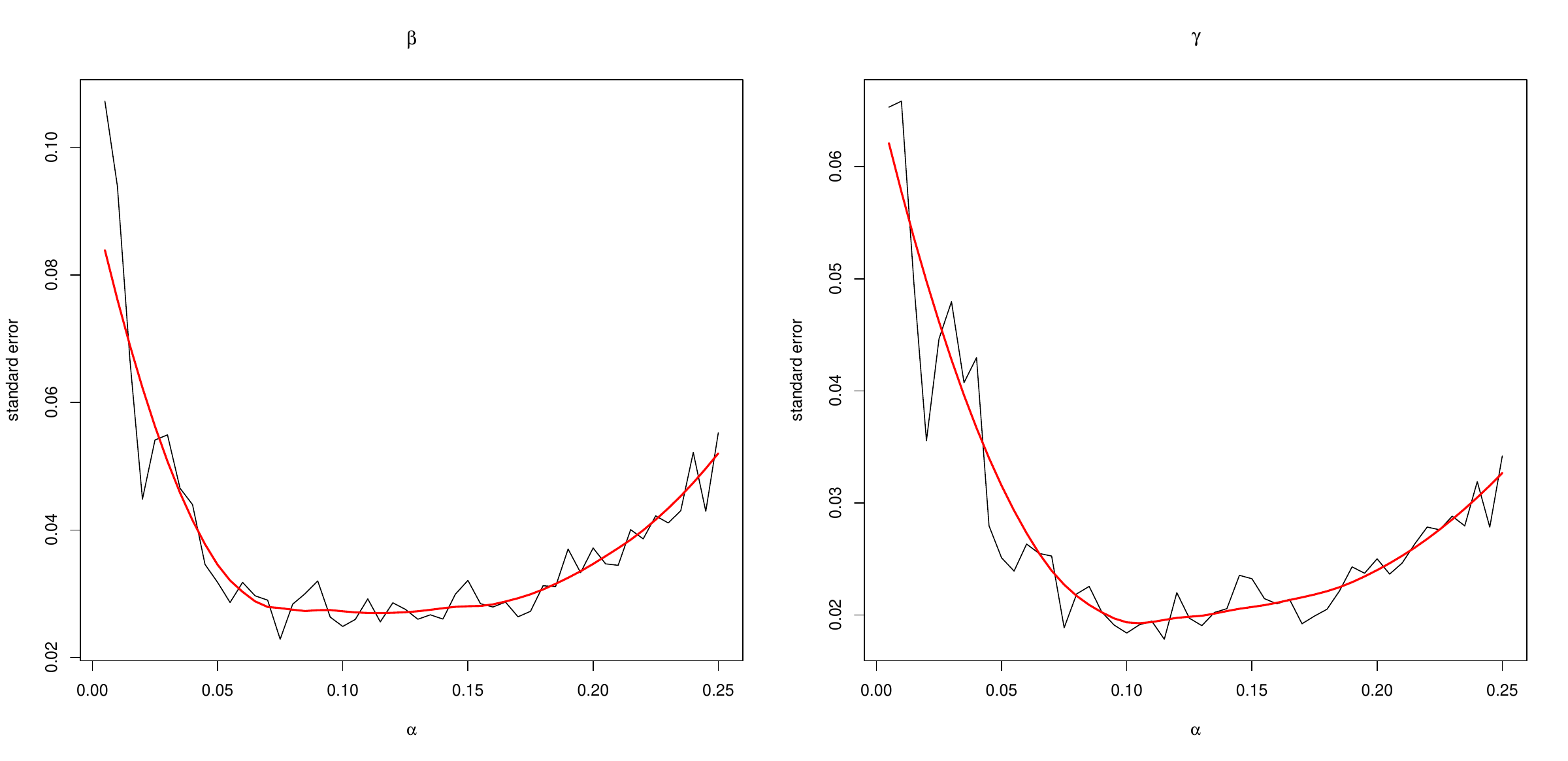}
\label{fig:asset_boot}
\end{figure}

Figures \ref{fig:index_boot}, for the three markets ($ASX\,200$, $FTSE\,100$, $S\&P\,500$), and \ref{fig:asset_boot}, for the three assets, ($A$, $GEN$, $MSFT$), using the full sample data from 2006-07-01 to 2023-09-30 and $B=500$ replications, plot the estimated bootstrap standard errors for $\hat{\omega}_{\alpha_{\text{est}}}$ and $\hat{\gamma}_{\alpha_{\text{est}}}$ against $\alpha_{\text{est}}$, the value of the quantile level used for estimation, which varies over the interval $0<\alpha_{\text{est}}\leq 0.25$ in steps of 0.005. Also, in order to facilitate the interpretation of the plots, a loess-smoothed curve has been added. In general, the empirical findings on the real data are qualitatively consistent with the evidence provided by the simulation study presented in Section \ref{ss:secaviar}. Although there is some variability across assets, the bootstrap standard errors tend to be higher when $\alpha_{\text{est}}$ is close to 0 or the upper bound of the grid 0.25.

Table \ref{tab:real_boot_raw} (top panel) provides additional information reporting the optimal value of $\alpha_{\text{est}}$ that, in most cases, tends to take values falling in the interval $[0.05,0.15]$ and, in any case, well above the target level $\alpha_0$. The bottom panel of the same table provides a robustness check against the impact of estimation noise by repeating the same exercise on the loess-smoothed curves (red lines in the plots). 

\begin{table}[!h]
\centering
\caption{Bootstrap results based on real data and 500 replications indicating the level of $\alpha$ that minimizes standard errors  for the estimation of $\beta$ and $\gamma$ (top: raw data; bottom: loess smoothed values). \\}
\label{tab:real_boot_raw}
\begin{tabular}{lcccccc}
\hline
                     & ASX200 & FTSE100 & SP500  & A      & GEN    & MSFT   \\ \hline
$\beta$              & 0.1350 & 0.0800 & 0.0950 & 0.1250 & 0.0550 & 0.0750 \\
$\gamma$             & 0.1350 & 0.1950 & 0.1300 & 0.1250 & 0.1650 & 0.1150 \\ \hline
$mean(\beta,\gamma)$ & 0.1350 & 0.1375 & 0.1125 & 0.1250 & 0.1100 & 0.0950 \\ \hline
$\beta$              & 0.1400 & 0.0650 & 0.1000 & 0.0950 & 0.0500 & 0.1150 \\
$\gamma$             & 0.1400 & 0.1950 & 0.1000 & 0.1450 & 0.1600 & 0.1050 \\ \hline
$mean(\beta,\gamma)$ & 0.1400 & 0.1300 & 0.1000 & 0.1200 & 0.1050 & 0.1100 \\ \hline
\end{tabular}%
\end{table}




\vspace{0.5cm}
{\centering \subsection{\normalsize Tail risk forecasting} \label{ss:riskfor} }
\noindent
This section presents the results from forecasting 1-day ahead and 10-day ahead VaR and ES levels, for the returns from the three assets MSFT, GEN and A and the three market indices SP500 (US), FTSE100 (UK) and ASX200 (Australia). Forecasts for VaR and ES are produced targeting risk levels $\alpha_0=0.025, 0.01$ from 28 competing models and methods. We consider filtered HS methods employing GARCH(1,1) and GJR-GARCH(1,1) as volatility equations, whilst filtering via Realized GARCH \citep{hansen2012realized} and Realized EGARCH \citep{hansen2016exponential} models are also included; all estimated by QML assuming Gaussian errors. The Realized GARCH and Realized EGARCH specifications employed are, respectively:
\begin{align*}
    \log \sigma_t^2 &= \alpha_0 + \alpha_1 \log X_{t-1}^2 + \beta_1 \log \sigma_{t-1}^2 \\ 
    \log X_t^2 &= \xi + \phi \log \sigma_t^2 + \tau_1 z_t + \tau_2 (z_t^2-1) + u_t
\end{align*}
\begin{align*}
    \log \sigma_t^2 &= \alpha_0 + \beta_1 \log \sigma_{t-1}^2 + \tau_1 z_{t-1} + \tau_2 (z_{t-1}^2-1) + \gamma u_{t-1} \\ 
    \log X_t^2 &= \xi + \phi \log \sigma_t^2 + \delta_1 z_t + \delta_2 (z_t^2-1) + u_t
\end{align*}
where in each case the measurement error follows $u_t \sim \text{i.i.d.} \, N(0,\tau^2)$ and the intra-day range (R) is used as realized measure (RM) $X_t$ in each model. 

The proposed quantile HS methods are included using quantile equations: CAViaR-IG (IG), CAViaR-IGJR (IGJR), Realized CAViaR (Re-C) and log Realized CAViaR (log Re-C). These four models map one-one to each of the GARCH, GJR-GARCH, Realized GARCH and Realized EGARCH volatility models employed, assuming that $\var_{\alpha,t} = c_{\alpha} \sigma_t$ in each case, and are estimated with $\alpha_{\text{est}} = 0.01, 0.025, 0.05, 0.1, 0.15, 0.2$. The Re-C and log Re-C \citep{peiris2024} specifications are, respectively:
\begin{align}
     \log |\q_{\alpha,t}| &= \alpha_0 + \alpha_1 \log X_{t-1} + \beta_1 \log |\q_{\alpha,t-1}| \\ 
    \log X_t &= \xi + \phi \log |\q_{\alpha,t}| + \tau_1 \epsilon_t + \tau_2 \epsilon_t^2 + u_t  \nonumber
\end{align}
\begin{align}
    \log |\q_{\alpha,t}| &= \alpha_0 + \beta_1 \log |\q_{\alpha,t-1}| + \tau_1 \epsilon_{t-1} + \tau_2 \epsilon_{t-1}^2 + \gamma u_{t-1} \\ 
    \log X_t &= \xi + \phi \log |\q_{\alpha,t}| + \delta_1 \epsilon_t + \delta_2 \epsilon_t^2 + u_t  \nonumber
\end{align}
where $\epsilon_{\alpha,t} = \frac{r_{t}}{-\q_{\alpha,t}}$ and the measurement error and RM are defined as above. Estimation for both models is done by maximising the pseudo-likelihoods, which combine the exponent of the negative quantile loss function with the Gaussian pdf from each measurement equation, as developed by \cite{gerlach_wang_2020} and \cite{peiris2024} for these models, respectively. The latter log Re-C model's parameters were optimised by running the Markov chain Monte Carlo algorithm in \cite{peiris2024} and taking the sampled parameter vector that maximised the pseudo-likelihood. For $h=1$ all models were re-estimated every $5$ days, whilst for $h=10$, all models were re-estimated every two 10 day periods, in the forecast period. For the sake of brevity, we report only aggregate results for the three assets and three markets of interest in the main text. Additional empirical results on the individual assets and markets are available on request. In the tables, we use the notational convention '$Cx$' to indicate that the semi-parametric model $C$ was estimated at level $x=100\alpha_{\text{est}}$. 



Table \ref{tab:allassets1day} presents a summary for the $h=1$ VaR and ES forecasts targeting $\alpha_0=0.025, 0.01$ for the three assets. Table \ref{tab:allmarkets1day} presents the same summary for the three markets. Included for each model/method are the quantile and joint loss values in the forecast period, averaged over the three assets, as well as the average ranks for each and the average MCS p-value for each model. The number of inclusions for each model in the MCS over the 3 assets is given in brackets below each MCS average p-value, whilst the final column adds these inclusions (which has a maximum value of 12 for each model: 3 assets, 4 loss values). 

First, it is immediately apparent that the GARCH-FHS (GHS) and GJR-GARCH-FHS (GJ-HS) methods are the least accurate VaR and ES forecasters for $h=1$ and $\alpha_0=0.025, 0.01$ for these three assets. Bolded numbers indicate the worst performing model for each metric, and these dominate the row for the GJR-GARCH-FHS methods. These two FHS methods are also included in the lowest number (4) of MCSs overall, compared to 6-9 inclusions for the CAViaR-IG (IG) and CAViaR-IGJR (IGJR) methods, i.e. the two QFHS methods are statistically more accurate than the two FHS methods here, especially for 1\% VaR and joint VaR, ES forecasting. Further, the QFHS IG and IGJR methods all have average quantile and joint loss, and average ranks, at both $\alpha_0 = 0.01, 0.025$, lower than their GHS and GJ-HS competitors; the QFHS methods also all have higher number of MCS inclusions and mostly higher average MCS p-values. 

Second, clearly the intra-day range has more efficient information than the daily returns for these assets, allowing all the Realized methods to outperform those only employing daily returns, as first evidenced by these models having typically more MCS inclusions. Boxes indicate the favoured model for each metric and these dominate the row for the log Re-C  method estimated with $\alpha_{\text{est}} = 0.2$; whilst the Re-GARCH (Re-GHS) method has the maximum of 12 MCS inclusions and is favoured, though not statistically, especially for joint $1\%$ VaR and ES joint forecasting. If pressed to choose a ``best'' model, it would be the log Re-C with $\alpha_{\text{est}} = 0.2$ for 2.5\% VaR and ES and also 1\% VaR forecasting (if done by itself), and the Re-GHS for 1\% joint VaR and ES forecasting. 

Table \ref{tab:allmarkets1day} is similar to Table \ref{tab:allassets1day}
 except that it presents the same metric averaged or cumulated over the forecast model results for the three markets. The results are quite different to those for the individual assets. First, the GHS and GJ-HS methods now perform very similarly to the QFHS IG and IGJR methods, respectively, with the latter now having varying performance over changing $\alpha_{\text{est}}$, with generally equivalent performance (compared to FHS) for $\alpha_{\text{est}} \le 0.1$, but with the least favourable results overall occurring when $\alpha_{\text{est}} \ge 0.15$. There seems to be no advantage for QFHS over FHS, for GARCH-type models, for these three market return series; while the GJR-based methods marginally outperform the GARCH-based methods in each case. 

Regarding the QFHS Realized models performance, the Re-EGARCH (Re-EGHS) and log Re-C methods significantly outperform the Re-GHS and Re-C methods, which performed very similarly to the GJ-HS method, e.g. having 9 MCS inclusions and very similar average loss values and ranks. The most clearly favoured models over the average losses and ranks are the log Re-C model with $\alpha_{\text{est}} \ge 0.1$, which have very (relatively and absolutely) low average ranks, especially for VaR and joint VaR, ES forecasting at $\alpha_0=0.01$. However, as the Re-EGHS and log Re-C methods all achieve the maximum of 12 MCS inclusions, these differences are not statistically significant. Nevertheless, if pressed to suggest one 'best' model, it would be the QFHS log Re-C with $\alpha_{\text{est}} = 0.1$ for 2.5\% VaR and ES forecasting, and the log Re-C with $\alpha_{\text{est}} = 0.2$ for 1\% VaR and ES forecasting. However, there seems to be no statistical or clear advantage for QFHS over FHS, for Realized-type models, for these three market return series.   

Tables \ref{tab:allassets10day} and \ref{tab:allmarkets10day}
 present a summary of the $h=10$ VaR and ES forecasts, targeting $\alpha_0=0.025, 0.01$, for the three assets (\ref{tab:allassets10day}) and three markets (\ref{tab:allmarkets10day}). For 10 day ahead forecasting for the three assets, Table \ref{tab:allassets10day} shows that the GARCH-based FHS and CAViaR based QFHS methods perform very similarly, and also are not typically included in the MCSs. The Re-GHS and  Re-C methods also perform similarly and are only marginally improved over the GARCH-based FHS and CAViaR based QFHS methods, e.g. being included in a couple more MCSs. The Re-EGHS method performs more favourably here, e.g. it is included in 9 MCSs and has lower average losses and ranks. Finally, the log Re-C model performs similarly to the Re-EGHS when $\alpha_{\text{est}} \le 0.025, \alpha_{\text{est}} = 0.2$; but clearly out-performs all methods for $\alpha_{\text{est}} \in [0.05,0.15]$. This out-performance is seen in typically lowest average losses and ranks, highest average MCS p-values and is statistically significant via the highest MCS inclusions of 11 or 12.    

Table \ref{tab:allmarkets10day} shows that the $h=10$ day forecast results for the markets are again quite different to those for the individual assets. Again, the GHS and GJ-HS methods now perform very similarly to the QFHS IG and IGJR methods, respectively, with the QFHS methods now having similar performance over changing $\alpha_{\text{est}}$. There seems to be no advantage for QFHS over FHS, for GARCH-type models, for these market return series and no clear advantage for asymmetric models (GJ-HS, IGJR) over non-asymmetric models (GHS, IG). 

Regarding the filtered Realized models performance, the Re-EGHS and QFHS log Re-C methods perform very similarly to the Re-GHS and Re-C methods, e.g. having 9-10 MCS inclusions and very similar average loss values and ranks. In this case, the GARCH-based FHS and CAViaR based QFHS methods marginally outperform the filtered Realized models, with 11 compared to 9-10 MCS inclusions and marginally lowest average losses and ranks. The GJ-HS and IGJR-HS $\alpha_{\text{est}} = 0.025$ marginally, but not statistically, out-perform the other models. Once again, however, there seems to be no statistical or clear advantage for QFHS over FHS, for both GARCH and Realized-type models, for these three market return series for 10 day ahead forecasting.

In summary, the QFHS methods tend to significantly out-perform the traditional FHS methods for the individual asset series, but not for the three market return series, for both 1 day and 10 day ahead VaR and ES forecasting. For 1 day ahead, as well as 10 day ahead forecasting for the three assets, the Realized models tend to statistically outperform the non-Realized models; while for 10 day ahead forecasting for the three markets the non-Realized models marginally out-performed the Realized models. Finally, the log Re-C QFHS method significantly outperformed all other models, including the Re-EGHS, for 10 day forecasting for the three individual asset series.  

\begin{center}
\begin{table}[htb]
\caption{\small 1-day risk forecasting results: average values across all assets. Key to table: \emph{Q$\alpha$}; average quantile loss value at the $\alpha\%$ ; \emph{J$\alpha$}; average joint loss value at the $\alpha\%$; \emph{R(L$\alpha$)}: average ranking of models under loss L at level $\alpha$; \emph{M(L$\alpha$)}: MCS p-value under loss L$\alpha$ (in brackets: number of times a model is included in the MCS); \emph{M(all)}: number of times a model is included in the MCS overall.\\}
\label{tab:allassets1day}
\setlength{\tabcolsep}{1pt}
\renewcommand{\arraystretch}{0.65}

\resizebox{\textwidth}{!}{%
\begin{tabular}{cccccccccccccc}
\hline
Model   & Q2.5   & J2.5  & Q1     & J1    & R(Q2.5) & R(J2.5) & R(Q1)  & R(J1)  & M(Q2.5)        & M(J2.5)        & M(Q1)          & M(J1)          & M(all) \\ \hline
GHS     & 218.18 & 2.78  & 120.37 & 3.14  & 23.3    & 19.8    & 23.3   & 16.7   & $\underset{(0)}{0.05}$ & $\underset{(1)}{0.37}$ & $\underset{(1)}{0.29}$ & $\underset{(2)}{0.65}$ & \textbf{4} \\
GJ-HS   & \textbf{220.96} & \textbf{2.79} & \textbf{121.91} & 3.16 & \textbf{26.3} & \textbf{24.3} & \textbf{26.0} & \textbf{22.2} & $\underset{(0)}{0.06}$ & $\underset{(1)}{0.34}$ & $\underset{(1)}{\textbf{0.26}}$ & $\underset{(2)}{0.62}$ & \textbf{4} \\ \hline
IG1     & 215.04 & 2.76  & 117.55 & 3.11  & 16.3    & 14.5    & 18.3   & 15.0   & $\underset{(0)}{0.14}$ & $\underset{(1)}{0.41}$ & $\underset{(3)}{0.60}$ & $\underset{(3)}{0.68}$ & 7 \\
2.5     & 215.35 & 2.77  & 116.83 & 3.10  & 17.7    & 17.5    & 13.7   & 13.3   & $\underset{(1)}{0.21}$ & $\underset{(1)}{0.37}$ & $\underset{(3)}{0.59}$ & $\underset{(3)}{0.67}$ & 8 \\
5       & 217.49 & 2.78  & 117.58 & 3.11  & 23.0    & 21.8    & 16.7   & 15.7   & $\underset{(0)}{0.05}$ & $\underset{(1)}{0.36}$ & $\underset{(3)}{0.52}$ & $\underset{(3)}{0.65}$ & 7 \\
10      & 217.69 & 2.78  & 118.16 & 3.14  & 22.7    & 22.3    & 18.3   & 19.0   & $\underset{(0)}{0.05}$ & $\underset{(1)}{0.37}$ & $\underset{(3)}{0.46}$ & $\underset{(2)}{0.66}$ & 6 \\
15      & 217.13 & 2.77  & 118.61 & 3.13  & 20.3    & 18.5    & 21.0   & 16.0   & $\underset{(0)}{0.10}$ & $\underset{(2)}{0.48}$ & $\underset{(2)}{0.47}$ & $\underset{(3)}{0.74}$ & 7 \\
20      & 216.85 & 2.77  & 118.75 & 3.11  & 20.7    & 19.0    & 21.0   & 17.0   & $\underset{(0)}{0.09}$ & $\underset{(1)}{0.38}$ & $\underset{(3)}{0.44}$ & $\underset{(3)}{0.66}$ & 7 \\
IGJR1   & 216.89 & 2.78  & 118.45 & 3.14  & 19.7    & 19.3    & 17.3   & 19.3   & $\underset{(1)}{0.23}$ & $\underset{(2)}{\textbf{0.25}}$ & $\underset{(3)}{0.58}$ & $\underset{(3)}{0.61}$ & 9 \\
2.5     & 216.19 & 2.77  & 118.31 & 3.12  & 19.0    & 20.7    & 19.0   & 19.8   & $\underset{(0)}{0.10}$ & $\underset{(1)}{0.37}$ & $\underset{(3)}{0.47}$ & $\underset{(3)}{0.63}$ & 7 \\
5       & 217.74 & 2.77  & 117.82 & 3.12  & 23.7    & 20.3    & 18.0   & 16.3   & $\underset{(0)}{0.07}$ & $\underset{(1)}{0.36}$ & $\underset{(3)}{0.49}$ & $\underset{(3)}{0.62}$ & 7 \\
10      & 216.10 & 2.76  & 116.79 & 3.11  & 19.7    & 17.3    & 13.3   & 13.0   & $\underset{(0)}{0.14}$ & $\underset{(2)}{0.54}$ & $\underset{(3)}{0.59}$ & $\underset{(3)}{0.77}$ & 8 \\
15      & 218.53 & 2.78  & 118.40 & 3.13  & 24.7    & 22.5    & 20.3   & 18.8   & $\underset{(0)}{0.07}$ & $\underset{(2)}{0.42}$ & $\underset{(2)}{0.46}$ & $\underset{(3)}{0.69}$ & 7 \\
20      & 217.45 & 2.77  & 118.76 & 3.12  & 24.0    & 20.7    & 21.3   & 18.7   & $\underset{(0)}{0.09}$ & $\underset{(1)}{0.38}$ & $\underset{(3)}{0.42}$ & $\underset{(3)}{0.62}$ & 7 \\ \hline
Re-GHS  & 207.98 & {\color{blue}2.712} & {\color{blue}114.13} & {\fbox{3.088}} & 6.3 & 4.5 & 8.3 & {\fbox{7.8}} & $\underset{(3)}{\fbox{0.88}}$ & $\underset{(3)}{\cb{0.82}}$ & $\underset{(3)}{0.80}$ & $\underset{(3)}{0.93}$ & {\fbox{12}} \\
Re-EGHS & 208.11 & 2.714  & 114.52 & 3.10  & 5.5     & {\color{blue}3.7} & 6.0 & {\fbox{7.8}} & $\underset{(2)}{0.57}$ & $\underset{(3)}{0.80}$ & $\underset{(3)}{0.80}$ & $\underset{(3)}{\cb{0.94}}$ & {\color{blue}11} \\ \hline
Re-C1   & 208.44 & 2.72  & 116.60 & 3.12  & 8.2     & 8.7     & 13.3   & 13.3   & $\underset{(2)}{0.42}$ & $\underset{(2)}{0.67}$ & $\underset{(3)}{0.66}$ & $\underset{(3)}{0.79}$ & 10 \\
2.5     & 209.18 & 2.74  & 118.56 & 3.15  & 11.5    & 16.2    & 14.0   & 13.8   & $\underset{(1)}{0.25}$ & $\underset{(2)}{0.53}$ & $\underset{(2)}{0.56}$ & $\underset{(3)}{0.56}$ & 8 \\
5       & 208.57 & 2.73  & 116.64 & 3.15  & 9.2     & 14.8    & 13.7   & 15.0   & $\underset{(2)}{0.40}$ & $\underset{(3)}{0.70}$ & $\underset{(3)}{0.71}$ & $\underset{(3)}{0.62}$ & {\color{blue}11} \\
10      & 208.82 & 2.74  & 121.52 & 3.16  & 9.8     & 16.7    & 16.3   & 18.2   & $\underset{(2)}{0.32}$ & $\underset{(2)}{0.48}$ & $\underset{(2)}{0.41}$ & $\underset{(2)}{\textbf{0.42}}$ & 8 \\
15      & 208.27 & 2.75  & 120.96 & \textbf{3.20} & 7.8 & 14.7 & 12.0 & 12.7 & $\underset{(2)}{0.46}$ & $\underset{(2)}{0.52}$ & $\underset{(2)}{0.68}$ & $\underset{(2)}{0.53}$ & 8 \\
20      & 209.73 & 2.72  & 117.21 & 3.12  & 10.7    & 11.3    & 9.7    & 10.2   & $\underset{(2)}{0.46}$ & $\underset{(3)}{0.76}$ & $\underset{(2)}{0.73}$ & $\underset{(3)}{0.89}$ & 10 \\ \hline
Re-EC1  & {\color{blue}207.73} & 2.72 & {\fbox{113.32}} & {\color{blue}3.095} & {\fbox{4.5}} & 6.0 & {\color{blue}5.7} & 8.2 & $\underset{(3)}{\cb{0.78}}$ & $\underset{(3)}{0.73}$ & $\underset{(3)}{\cb{0.86}}$ & $\underset{(3)}{0.79}$ & {\fbox{12}} \\
2.5     & 208.36 & 2.72  & 114.44 & 3.10  & 6.2     & 6.7     & 7.0    & 11.7   & $\underset{(2)}{0.45}$ & $\underset{(2)}{0.69}$ & $\underset{(3)}{0.76}$ & $\underset{(3)}{0.81}$ & 10 \\
5       & 207.83 & 2.72  & 114.86 & 3.10  & 6.2     & 5.5     & 9.0    & 12.2   & $\underset{(2)}{0.54}$ & $\underset{(3)}{0.78}$ & $\underset{(3)}{0.76}$ & $\underset{(3)}{0.87}$ & {\color{blue}11} \\
10      & 208.46 & 2.72  & 114.95 & 3.11  & 8.2     & 8.5     & 10.3   & 13.3   & $\underset{(2)}{0.46}$ & $\underset{(3)}{0.73}$ & $\underset{(3)}{0.73}$ & $\underset{(3)}{0.78}$ & {\color{blue}11} \\
15      & 208.02 & 2.72  & 114.50 & 3.12  & 5.8     & 7.8     & 8.3    & 13.0   & $\underset{(2)}{0.49}$ & $\underset{(3)}{0.75}$ & $\underset{(3)}{0.85}$ & $\underset{(3)}{0.76}$ & {\color{blue}11} \\
20      & {\fbox{207.72}} & {\fbox{2.711}} & 114.24 & {\fbox{3.088}} & {\color{blue}5.2} & {\fbox{2.3}} & {\fbox{4.7}} & {\color{blue}8.0} & $\underset{(2)}{0.61}$ & $\underset{(3)}{\fbox{0.96}}$ & $\underset{(3)}{\fbox{0.87}}$ & $\underset{(3)}{\fbox{1.00}}$ & {\color{blue}11} \\ \hline
\end{tabular}
}
\end{table}
\end{center}

\begin{center}
 \begin{table}[htb]
\caption{\small 1-day risk forecasting results: average values across all markets. Key to table: \emph{Q$\alpha$}; average quantile loss value at the $\alpha\%$ ; \emph{J$\alpha$}; average joint loss value at the $\alpha\%$; \emph{R(L$\alpha$)}: average ranking of models under loss L at level $\alpha$; \emph{M(L$\alpha$)}: MCS p-value under loss L$\alpha$ (in brackets: number of times a model is included in the MCS); \emph{M(all)}: number of times a model is included in the MCS overall. \\}
\label{tab:allmarkets1day}
\setlength{\tabcolsep}{1pt}
\renewcommand{\arraystretch}{0.65}

\resizebox{\textwidth}{!}{%
\begin{tabular}{cccccccccccccc}
\hline
Model   & Q2.5                          & J2.5                        & Q1                           & J1                          & R(Q2.5)                   & R(J2.5)                   & R(Q1)                     & R(J1)                     & M(Q2.5)                          & M(J2.5)                         & M(Q1)                            & M(J1)                            & M(all)                    \\ \hline
{GHS}     & 113.35                         & 2.06                        & 56.12                        & 2.28                        & 15.7                      & 19.2                      & 15.3                      & 19.0                      & $\underset{(2)}{0.36}$           & $\underset{(1)}{0.25}$          & $\underset{(2)}{0.45}$           & $\underset{(2)}{0.41}$           & 7                         \\
GJ-HS   & 110.24                         & 2.03                        & 54.97                        & 2.27                        & 6.7                       & 10.8                      & 11.7                      & 17.3                      & $\underset{(3)}{0.82}$           & $\underset{(1)}{0.47}$          & $\underset{(3)}{0.82}$           & $\underset{(2)}{0.51}$           & 9                         \\ \hline
IG1     & 112.48                         & 2.05                        & 55.09                        & 2.25                        & 14.7                      & 16.3                      & 11.3                      & 16.8                      & $\underset{(2)}{0.50}$           & $\underset{(1)}{0.40}$          & $\underset{(2)}{0.69}$           & $\underset{(2)}{0.62}$           & 7                         \\
2.5     & 113.09                         & 2.06                        & 55.50                        & 2.25                        & 17.0                      & 18.7                      & 14.0                      & 17.3                      & $\underset{(2)}{0.45}$           & $\underset{(1)}{0.39}$          & $\underset{(2)}{0.68}$           & $\underset{(2)}{0.58}$           & 7                         \\
5       & 113.41                         & 2.06                        & 56.49                        & 2.28                        & 17.3                      & 20.3                      & 17.3                      & 19.7                      & $\underset{(2)}{0.35}$           & $\underset{(1)}{0.29}$          & $\underset{(2)}{0.41}$           & $\underset{(2)}{0.50}$           & 7                         \\
10      & 114.23                         & 2.07                        & 56.66                        & 2.29                        & 18.0                      & 19.7                      & 14.3                      & 18.2                      & $\underset{(2)}{0.29}$           & $\underset{(1)}{0.17}$          & $\underset{(1)}{0.38}$           & $\underset{(1)}{0.32}$           & 5                         \\
15      & 116.35                         & 2.09                        & 58.62                        & 2.31                        & 26.3                      & 22.3                      & \textbf{25.3}             & 21.3                      & $\underset{(1)}{\textbf{0.21}}$  & $\underset{(1)}{\textbf{0.13}}$  & $\underset{(1)}{\textbf{0.24}}$  & $\underset{(1)}{0.25}$           & \textbf{4}                \\
20      & \textbf{116.85}                & \textbf{2.09}               & \textbf{59.44}               & \textbf{2.33}               & \textbf{26.7}             & \textbf{24.2}             & \textbf{25.3}             & \textbf{24.3}             & $\underset{(2)}{0.25}$           & $\underset{(1)}{0.14}$          & $\underset{(2)}{0.33}$           & $\underset{(1)}{\textbf{0.22}}$  & 6                         \\
IGJR1   & 113.54                         & 2.06                        & 55.91                        & 2.28                        & 18.7                      & 19.8                      & 15.0                      & 19.7                      & $\underset{(2)}{0.42}$           & $\underset{(1)}{0.37}$          & $\underset{(2)}{0.59}$           & $\underset{(2)}{0.56}$           & 7                         \\
2.5     & 112.79                         & 2.07                        & 56.37                        & 2.29                        & 15.3                      & 19.5                      & 17.0                      & 19.0                      & $\underset{(2)}{0.51}$           & $\underset{(1)}{0.36}$          & $\underset{(2)}{0.53}$           & $\underset{(2)}{0.44}$           & 7                         \\
5       & 112.16                         & 2.05                        & 56.68                        & 2.30                        & 13.2                      & 16.5                      & 19.3                      & 19.7                      & $\underset{(3)}{0.52}$           & $\underset{(1)}{0.37}$          & $\underset{(2)}{0.45}$           & $\underset{(1)}{0.44}$           & 7                         \\
10      & 111.10                         & 2.04                        & 55.28                        & 2.27                        & 10.3                      & 16.8                      & 13.3                      & 17.3                      & $\underset{(3)}{0.72}$           & $\underset{(1)}{0.41}$          & $\underset{(3)}{0.71}$           & $\underset{(2)}{0.42}$           & 9                         \\
15      & 113.86                         & 2.07                        & 56.87                        & 2.31                        & 19.7                      & 17.3                      & 18.0                      & 17.7                      & $\underset{(3)}{0.52}$           & $\underset{(2)}{0.33}$          & $\underset{(2)}{0.50}$           & $\underset{(2)}{0.37}$           & 9                         \\
20      & 115.05                         & 2.07                        & 57.71                        & 2.32                        & 23.3                      & 20.3                      & 21.0                      & 20.8                      & $\underset{(2)}{0.32}$           & $\underset{(1)}{0.21}$          & $\underset{(3)}{0.51}$           & $\underset{(1)}{0.26}$           & 7                         \\ \hline
Re-GHS  & 113.88                         & 2.05                        & 57.17                        & 2.28                        & 18.2                      & 16.3                      & 17.7                      & 15.3                      & $\underset{(2)}{0.41}$           & $\underset{(2)}{0.30}$          & $\underset{(3)}{0.46}$           & $\underset{(2)}{0.41}$           & 9                         \\
Re-EGHS & 109.97                         & 2.00                        & 54.44                        & 2.20                        & 7.0                       & 6.3                       & 9.3                       & 7.8                       & $\underset{(3)}{0.80}$           & $\underset{(3)}{0.68}$          & $\underset{(3)}{0.82}$           & $\underset{(3)}{0.80}$           & {\fbox{12}} \\ \hline
Re-C1   & 114.14                         & 2.05                        & 57.20                        & 2.27                        & 19.7                      & 17.5                      & 18.3                      & 14.3                      & $\underset{(2)}{0.35}$           & $\underset{(2)}{0.26}$          & $\underset{(3)}{0.43}$           & $\underset{(1)}{0.39}$           & 8                         \\
2.5     & 114.30                         & 2.05                        & 57.38                        & 2.27                        & 19.0                      & 14.8                      & 20.0                      & 13.8                      & $\underset{(3)}{0.50}$           & $\underset{(1)}{0.32}$          & $\underset{(3)}{0.59}$           & $\underset{(3)}{0.50}$           & {\color{blue}10} \\
5       & 113.67                         & 2.04                        & 56.81                        & 2.27                        & 17.3                      & 15.3                      & 16.3                      & 14.0                      & $\underset{(2)}{0.46}$           & $\underset{(2)}{0.43}$          & $\underset{(3)}{0.59}$           & $\underset{(2)}{0.53}$           & 9                         \\
10      & 113.93                         & 2.04                        & 57.28                        & 2.28                        & 18.3                      & 15.5                      & 19.0                      & 14.2                      & $\underset{(3)}{0.41}$           & $\underset{(2)}{0.28}$          & $\underset{(3)}{0.46}$           & $\underset{(1)}{0.39}$           & 9                         \\
15      & 114.37                         & 2.05                        & 57.38                        & 2.28                        & 21.7                      & 18.3                      & 20.7                      & 14.8                      & $\underset{(2)}{0.39}$           & $\underset{(2)}{0.28}$          & $\underset{(3)}{0.48}$           & $\underset{(1)}{0.39}$           & 8                         \\
20      & 113.95                         & 2.04                        & 57.13                        & 2.27                        & 18.3                      & 16.0                      & 19.3                      & 13.3                      & $\underset{(3)}{0.47}$           & $\underset{(2)}{0.32}$          & $\underset{(3)}{0.53}$           & $\underset{(1)}{0.44}$           & 9                         \\ \hline
Re-EC1  & 109.82                         & 2.00                        & 54.31                        & 2.21                        & 6.0                       & 5.2                       & 8.0                       & 8.7                       & $\underset{(3)}{0.80}$           & $\underset{(3)}{0.68}$          & $\underset{(3)}{0.89}$           & $\underset{(3)}{0.82}$           & {\fbox{12}} \\
2.5     & 109.42                         & 2.00                        & 53.63                        & 2.20                        & {\fbox{3.0}} & 4.2                       & 6.0                       & 7.8                       & $\underset{(3)}{0.80}$           & $\underset{(3)}{0.69}$          & $\underset{(3)}{0.95}$           & $\underset{(3)}{0.90}$           & {\fbox{12}} \\
5       & 109.09                         & 1.994                        & 53.57                        & 2.20                        & 4.0                       & 4.2                       & 5.3                       & 6.7                       & $\underset{(3)}{0.83}$           & $\underset{(3)}{0.68}$          & $\underset{(3)}{0.92}$           & $\underset{(3)}{0.84}$           & {\fbox{12}} \\
10      & {\fbox{108.97}} & {\fbox{1.989}} & {\color{blue}53.31} & {\color{blue}2.187} & 4.0                       & {\fbox{3.2}} & {\color{blue}2.3} & {\color{blue}2.0} & $\underset{(3)}{\fbox{0.95}}$    & $\underset{(3)}{\fbox{0.89}}$   & $\underset{(3)}{\cb{0.96}}$      & $\underset{(3)}{\fbox{0.98}}$    & {\fbox{12}} \\
15      & {\color{blue}109.01} & {\color{blue}1.992} & 53.40                        & 2.19                        & {\color{blue}3.3} & {\color{blue}3.5} & 3.3                       & 3.7                       & $\underset{(3)}{\cb{0.87}}$      & $\underset{(3)}{\cb{0.79}}$     & $\underset{(3)}{\cb{0.96}}$      & $\underset{(3)}{0.90}$           & {\fbox{12}} \\
20      & 109.06                         & {\color{blue}1.992} & {\fbox{53.25}} & {\fbox{2.186}} & {\color{blue}3.3} & 3.8                       & {\fbox{2.0}} & {\fbox{1.3}} & $\underset{(3)}{0.83}$           & $\underset{(3)}{0.68}$          & $\underset{(3)}{\fbox{0.99}}$    & $\underset{(3)}{\cb{0.97}}$      & {\fbox{12}} \\ \hline
\end{tabular}
}
\end{table}
\end{center}

\begin{center}
 \begin{table}[hbt]
\caption{\small 10-day risk forecasting results: average values across all assets. Key to table: \emph{Q$\alpha$}; average quantile loss value at the $\alpha\%$ ; \emph{J$\alpha$}; average joint loss value at the $\alpha\%$; \emph{R(L$\alpha$)}: average ranking of models under loss L at level $\alpha$; \emph{M(L$\alpha$)}: MCS p-value under loss L$\alpha$ (in brackets: number of times a model is included in the MCS); \emph{M(all)}: number of times a model is included in the MCS overall.\\}
\label{tab:allassets10day}
\setlength{\tabcolsep}{1pt}
\renewcommand{\arraystretch}{0.65}
\resizebox{\textwidth}{!}{%
\begin{tabular}{cccccccccccccc}
\hline
Model   & Q2.5     & J2.5    & Q1      & J1      & R(Q2.5) & R(J2.5) & R(Q1)   & R(J1)   & M(Q2.5)                    & M(J2.5)                    & M(Q1)                      & M(J1)                      & M(all) \\ \hline
GHS     & \textbf{155.62} & 3.87    & 72.66   & 4.00    & \textbf{25.7} & \textbf{23.3} & 17.7    & 19.3    & $\underset{(1)}{0.16}$     & $\underset{(0)}{0.07}$     & $\underset{(0)}{0.03}$     & $\underset{(1)}{0.22}$     & 2      \\
GJ-HS   & 152.85   & 3.83    & 69.88   & 3.93    & 22.7    & 16.5    & 10.7    & 9.8     & $\underset{(2)}{0.28}$     & $\underset{(1)}{0.26}$     & $\underset{(0)}{0.11}$     & $\underset{(1)}{0.32}$     & 4      \\ \hline
IG1     & 154.10   & 3.84    & 70.31   & 3.98    & 20.0    & 17.3    & 11.3    & 15.7    & $\underset{(1)}{0.30}$     & $\underset{(1)}{0.33}$     & $\underset{(0)}{0.02}$     & $\underset{(1)}{0.30}$     & 3      \\
2.5     & 150.96   & 3.82    & 71.35   & 3.97    & 16.3    & 15.5    & 15.0    & 16.7    & $\underset{(0)}{0.21}$     & $\underset{(0)}{0.10}$     & $\underset{(0)}{0.02}$     & $\underset{(1)}{0.26}$     & \textbf{1} \\
5       & 150.98   & 3.82    & 72.79   & 3.98    & 18.3    & 17.0    & 18.3    & 19.3    & $\underset{(2)}{0.24}$     & $\underset{(1)}{0.14}$     & $\underset{(0)}{\textbf{0.00}}$ & $\underset{(1)}{0.25}$     & 4      \\
10      & 153.65   & 3.84    & 73.26   & 3.99    & 23.7    & 20.7    & 19.0    & 17.8    & $\underset{(1)}{0.16}$     & $\underset{(0)}{0.07}$     & $\underset{(0)}{\textbf{0.00}}$ & $\underset{(0)}{0.10}$     & \textbf{1} \\
15      & 152.58   & 3.83    & 72.68   & 3.97    & 20.0    & 18.3    & 16.0    & 14.3    & $\underset{(0)}{0.17}$     & $\underset{(0)}{0.11}$     & $\underset{(0)}{\textbf{0.00}}$ & $\underset{(1)}{0.19}$     & \textbf{1} \\
20      & 152.93   & 3.83    & 72.68   & 3.97    & 21.3    & 18.0    & 17.0    & 14.2    & $\underset{(0)}{0.17}$     & $\underset{(0)}{0.12}$     & $\underset{(0)}{0.06}$     & $\underset{(1)}{0.16}$     & \textbf{1} \\
IGJR1   & 154.48   & 3.83    & 71.92   & 3.96    & 20.7    & 17.8    & 16.3    & 15.0    & $\underset{(0)}{\textbf{0.14}}$ & $\underset{(1)}{0.16}$     & $\underset{(0)}{0.05}$     & $\underset{(1)}{0.30}$     & 2      \\
2.5     & 151.80   & 3.83    & 72.98   & 3.99    & 17.7    & 15.5    & 19.0    & 19.7    & $\underset{(1)}{0.29}$     & $\underset{(1)}{0.31}$     & $\underset{(0)}{\textbf{0.00}}$ & $\underset{(1)}{0.26}$     & 3      \\
5       & 152.34   & 3.83    & 73.59   & 3.99    & 22.0    & 18.3    & 19.3    & 19.2    & $\underset{(1)}{0.15}$     & $\underset{(0)}{\textbf{0.06}}$ & $\underset{(0)}{\textbf{0.00}}$ & $\underset{(0)}{0.06}$     & \textbf{1} \\
10      & 151.06   & 3.82    & 74.44   & 4.00    & 18.0    & 16.3    & \textbf{21.3} & 20.2    & $\underset{(1)}{0.16}$     & $\underset{(0)}{0.08}$     & $\underset{(0)}{\textbf{0.00}}$ & $\underset{(0)}{\textbf{0.05}}$ & \textbf{1} \\
15      & 150.98   & 3.82    & 74.29   & 4.01    & 16.3    & 16.0    & 20.3    & 20.0    & $\underset{(1)}{0.19}$     & $\underset{(0)}{0.11}$     & $\underset{(0)}{\textbf{0.00}}$ & $\underset{(1)}{0.12}$     & 2      \\
20      & 150.37   & 3.81    & 73.99   & 3.99    & 15.3    & 14.2    & 20.3    & 15.7    & $\underset{(1)}{0.24}$     & $\underset{(1)}{0.22}$     & $\underset{(0)}{0.03}$     & $\underset{(0)}{0.10}$     & 2      \\ \hline
Re-GHS  & 147.45   & \textbf{4.00} & 70.98   & \textbf{4.21} & 9.7     & 18.2    & 10.5    & 14.8    & $\underset{(2)}{0.42}$     & $\underset{(2)}{0.46}$     & $\underset{(0)}{\textbf{0.00}}$ & $\underset{(2)}{0.38}$     & 6      \\
Re-EGHS & 143.24   & 3.74    & 70.98   & 3.94    & 7.0     & 6.8     & 10.5    & 10.0    & $\underset{(3)}{0.56}$     & $\underset{(3)}{0.64}$     & $\underset{(1)}{0.28}$     & $\underset{(2)}{0.37}$     & 9      \\ \hline
Re-C1   & 148.54   & 3.94    & 74.22   & 4.17    & 12.0    & 20.7    & 19.3    & \textbf{21.3} & $\underset{(3)}{0.53}$     & $\underset{(1)}{0.37}$     & $\underset{(0)}{\textbf{0.00}}$ & $\underset{(1)}{0.28}$     & 5      \\
2.5     & 149.57   & 3.82    & 73.78   & 3.98    & 14.7    & 15.8    & 13.7    & 12.0    & $\underset{(2)}{0.48}$     & $\underset{(1)}{0.39}$     & $\underset{(0)}{\textbf{0.00}}$ & $\underset{(1)}{0.37}$     & 4      \\
5       & 147.64   & 3.81    & 74.37   & 3.99    & 8.7     & 15.3    & 17.7    & 17.7    & $\underset{(2)}{0.49}$     & $\underset{(1)}{0.40}$     & $\underset{(1)}{0.22}$     & $\underset{(1)}{0.31}$     & 5      \\
10      & 148.48   & 3.89    & 72.11   & 4.09    & 12.3    & 18.8    & 16.0    & 19.0    & $\underset{(2)}{0.46}$     & $\underset{(1)}{0.37}$     & $\underset{(0)}{\textbf{0.00}}$ & $\underset{(1)}{0.29}$     & 4      \\
15      & 149.65   & 3.97    & 74.60   & 4.20    & 16.7    & 19.5    & 20.7    & 21.0    & $\underset{(2)}{0.47}$     & $\underset{(1)}{0.37}$     & $\underset{(0)}{\textbf{0.00}}$ & $\underset{(1)}{0.29}$     & 4      \\
20      & 149.17   & 3.97    & 74.95   & 4.21    & 14.7    & 20.8    & 20.0    & \textbf{21.3} & $\underset{(2)}{0.45}$     & $\underset{(1)}{0.37}$     & $\underset{(0)}{\textbf{0.00}}$ & $\underset{(1)}{0.29}$     & 4      \\ \hline
Re-EC1  & 145.64   & 3.77    & 68.12   & 3.90    & 8.0     & 6.0     & {\color{blue}4.3} & 4.3     & $\underset{(3)}{0.56}$     & $\underset{(2)}{0.57}$     & $\underset{(2)}{\cb{0.65}}$     & $\underset{(2)}{0.67}$     & 9      \\
2.5     & 143.07   & 3.74    & 66.12   & 3.87    & 4.7     & 5.7     & {\fbox{4.0}} & 4.2     & $\underset{(3)}{0.56}$     & $\underset{(2)}{0.64}$     & $\underset{(3)}{0.59}$     & $\underset{(2)}{0.67}$     & 10     \\
5       & {\color{blue}141.82} & 3.734    & 65.73   & 3.87    & {\fbox{3.0}} & {\color{blue}3.2} & {\fbox{4.0}} & 4.3     & $\underset{(3)}{0.68}$     & $\underset{(3)}{0.66}$     & $\underset{(3)}{0.61}$     & $\underset{(3)}{0.57}$     & {\fbox{12}} \\
10      & {\fbox{141.74}} & {\fbox{3.729}} & {\fbox{65.51}} & {\color{blue}3.86} & {\color{blue}4.0} & 3.5     & 4.7     & {\color{blue}4.0} & $\underset{(3)}{\fbox{0.79}}$   & $\underset{(3)}{\fbox{0.90}}$   & $\underset{(2)}{0.62}$     & $\underset{(3)}{\cb{0.71}}$     & {\color{blue}11} \\
15      & 142.28   & {\color{blue}3.730} & {\color{blue}65.71} & {\fbox{3.86}} & 5.0     & {\fbox{1.7}} & 5.7     & {\fbox{3.8}} & $\underset{(3)}{\cb{0.71}}$     & $\underset{(3)}{\cb{0.76}}$     & $\underset{(2)}{\fbox{0.67}}$     & $\underset{(3)}{\fbox{0.92}}$   & {\color{blue}11} \\
20      & 143.21   & 3.74    & 72.99   & 3.96    & 7.7     & 5.2     & 13.3    & 11.3    & $\underset{(3)}{0.46}$     & $\underset{(2)}{0.62}$     & $\underset{(1)}{0.20}$     & $\underset{(2)}{0.42}$     & 8      \\ \hline
\end{tabular}
}
\end{table}
\end{center}

\begin{center}
 \begin{table}[htb]
\caption{\small 10-day risk forecasting results: average values across all markets. Key to table: \emph{Q$\alpha$}; average quantile loss value at the $\alpha\%$ ; \emph{J$\alpha$}; average joint loss value at the $\alpha\%$; \emph{R(L$\alpha$)}: average ranking of models under loss L at level $\alpha$; \emph{M(L$\alpha$)}: MCS p-value under loss L$\alpha$ (in brackets: number of times a model is included in the MCS); \emph{M(all)}: number of times a model is included in the MCS overall.\\}
\label{tab:allmarkets10day}
\setlength{\tabcolsep}{1pt}
\renewcommand{\arraystretch}{0.65}

\resizebox{\textwidth}{!}{%
\begin{tabular}{cccccccccccccc}
\hline
Model   & Q2.5                          & J2.5                         & Q1                            & J1                           & R(Q2.5)                    & R(J2.5)                    & R(Q1)                      & R(J1)                      & M(Q2.5)                          & M(J2.5)                          & M(Q1)                            & M(J1)                           & M(all)                    \\ \hline
GHS     & 95.36                        & 3.36                        & 49.12                        & 3.62                        & 19.0                       & 24.5                       & 13.7                       & 20.0                       & $\underset{(3)}{0.52}$          & $\underset{(3)}{0.40}$          & $\underset{(2)}{0.41}$          & $\underset{(3)}{0.51}$         & {\fbox{11}} \\
GJ-HS   & {\color{blue}91.10} & {\fbox{3.252}} & 46.87                        & 3.49                        & {\fbox{4.3}} & {\fbox{4.7}} & {\color{blue}4.3} & 6.8                        & $\underset{(3)}{\fbox{0.86}}$   & $\underset{(3)}{\cb{0.81}}$     & $\underset{(2)}{\cb{0.60}}$     & $\underset{(3)}{0.70}$         & {\fbox{11}} \\ \hline
IG1     & 93.67                        & 3.29                        & {\color{blue}46.869} & 3.50                        & 13.0                       & 10.3                       & 7.0                        & 7.3                        & $\underset{(3)}{0.63}$          & $\underset{(3)}{0.50}$          & $\underset{(2)}{\fbox{0.66}}$   & $\underset{(3)}{0.75}$         & {\fbox{11}} \\
2.5     & 93.57                        & 3.29                        & 48.04                        & 3.53                        & 13.7                       & 10.2                       & 9.7                        & 10.7                       & $\underset{(3)}{0.62}$          & $\underset{(3)}{0.50}$          & $\underset{(2)}{0.42}$          & $\underset{(3)}{0.69}$         & {\fbox{11}} \\
5       & 93.49                        & 3.30                        & 48.41                        & 3.56                        & 11.7                       & 14.3                       & 12.7                       & 15.3                       & $\underset{(3)}{0.57}$          & $\underset{(3)}{0.48}$          & $\underset{(1)}{0.35}$           & $\underset{(3)}{0.60}$           & {\color{blue}10} \\
10      & 92.87                        & 3.31                        & 48.61                        & 3.59                        & 13.7                       & 17.0                       & 12.7                       & 19.0                       & $\underset{(3)}{0.60}$          & $\underset{(3)}{0.49}$          & $\underset{(1)}{0.35}$          & $\underset{(3)}{0.55}$         & {\color{blue}10} \\
15      & 92.06                        & 3.32                        & 49.96                        & 3.64                        & 12.3                       & 19.5                       & 14.7                       & 21.3                       & $\underset{(3)}{0.69}$          & $\underset{(3)}{0.48}$          & $\underset{(1)}{0.32}$          & $\underset{(3)}{0.50}$           & {\color{blue}10} \\
20      & 93.52                        & 3.33                        & 49.92                        & 3.64                        & 16.3                       & 21.0                       & 14.7                       & 24.0                       & $\underset{(3)}{0.53}$          & $\underset{(3)}{0.45}$          & $\underset{(1)}{0.34}$          & $\underset{(3)}{0.52}$         & {\color{blue}10} \\
IGJR1   & 94.76                        & 3.28                        & 47.45                        & {\color{blue}3.485} & 18.7                       & 11.2                       & 11.0                       & {\color{blue}6.3} & $\underset{(3)}{0.60}$            & $\underset{(3)}{0.49}$          & $\underset{(2)}{0.54}$          & $\underset{(3)}{\cb{0.76}}$    & {\fbox{11}} \\
2.5     & 92.82                        & 3.26                        & {\fbox{46.79}} & {\fbox{3.47}} & 14.7                       & 7.0                        & {\fbox{3.7}} & {\fbox{3.8}} & $\underset{(3)}{0.61}$          & $\underset{(3)}{0.69}$          & $\underset{(2)}{0.58}$          & $\underset{(3)}{\fbox{0.77}}$  & {\fbox{11}} \\
5       & 92.26                        & 3.27                        & 47.86                        & 3.51                        & 11.7                       & 8.7                        & 10.7                       & 9.3                        & $\underset{(3)}{0.62}$          & $\underset{(3)}{0.49}$          & $\underset{(1)}{0.31}$          & $\underset{(3)}{0.63}$         & {\color{blue}10} \\
10      & {\fbox{90.95}} & {\color{blue}3.258} & 47.71                        & 3.53                        & {\color{blue}5.7} & {\color{blue}5.5} & 10.0                       & 11.0                       & $\underset{(3)}{\cb{0.82}}$     & $\underset{(3)}{\fbox{0.82}}$   & $\underset{(2)}{0.58}$          & $\underset{(3)}{0.63}$         & {\fbox{11}} \\
15      & 91.84                        & 3.29                        & 49.09                        & 3.58                        & 7.3                        & 11.0                       & 12.7                       & 11.3                       & $\underset{(3)}{0.77}$          & $\underset{(3)}{0.66}$          & $\underset{(2)}{0.52}$          & $\underset{(3)}{0.69}$          & {\fbox{11}} \\
20      & 91.46                        & 3.27                        & 49.52                        & 3.56                        & 8.0                        & 9.3                        & 11.7                       & 12.3                       & $\underset{(3)}{0.77}$          & $\underset{(3)}{0.48}$           & $\underset{(1)}{0.29}$           & $\underset{(3)}{0.58}$         & {\color{blue}10} \\ \hline
Re-GHS  & 96.27                        & 3.36                        & 51.56                        & 3.68                        & 19.7                       & 20.5                       & 19.5                       & 18.7                       & $\underset{(3)}{0.53}$          & $\underset{(3)}{0.36}$          & $\underset{(0)}{\textbf{0.00}}$ & $\underset{(3)}{0.49}$         & 9                         \\
Re-EGHS & 94.85                        & 3.29                        & 51.56                        & 3.59                        & 11.3                       & 11.5                       & 19.5                       & 16.3                       & $\underset{(2)}{0.63}$          & $\underset{(3)}{0.64}$          & $\underset{(2)}{0.34}$          & $\underset{(3)}{0.43}$         & {\color{blue}10} \\ \hline
Re-C1   & 96.30                        & 3.34                        & 51.75                        & 3.65                        & 19.3                       & 18.0                       & 20.3                       & 17.3                       & $\underset{(3)}{0.48}$           & $\underset{(3)}{0.45}$          & $\underset{(0)}{0.06}$          & $\underset{(2)}{\textbf{0.40}}$ & \textbf{8}                \\
2.5     & 97.00                        & 3.37                        & 52.64                        & 3.72                        & 22.0                       & 23.3                       & \textbf{26.0}              & 23.0                       & $\underset{(2)}{0.41}$          & $\underset{(3)}{0.38}$          & $\underset{(1)}{0.11}$          & $\underset{(3)}{0.58}$         & 9                         \\
5       & 96.67                        & 3.37                        & 52.27                        & 3.71                        & 19.7                       & 21.7                       & 22.0                       & 22.3                       & $\underset{(3)}{0.43}$          & $\underset{(3)}{0.37}$          & $\underset{(1)}{0.13}$          & $\underset{(3)}{0.47}$         & {\color{blue}10} \\
10      & 96.10                        & 3.36                        & 51.50                        & 3.70                        & 19.0                       & 21.7                       & 20.3                       & 21.0                       & $\underset{(3)}{0.51}$          & $\underset{(3)}{0.33}$          & $\underset{(1)}{0.17}$          & $\underset{(3)}{0.50}$         & {\color{blue}10} \\
15      & 96.94                        & 3.37                        & 52.61                        & 3.72                        & 22.3                       & 23.5                       & 25.3                       & 24.0                       & $\underset{(3)}{0.37}$          & $\underset{(3)}{0.30}$          & $\underset{(0)}{\textbf{0.00}}$ & $\underset{(3)}{0.42}$         & 9                         \\
20      & \textbf{97.49}               & \textbf{3.38}               & \textbf{53.21}               & \textbf{3.74}               & \textbf{24.7}              & \textbf{26.2}              & 25.7                       & \textbf{24.3}              & $\underset{(2)}{\textbf{0.30}}$ & $\underset{(3)}{\textbf{0.29}}$ & $\underset{(0)}{\textbf{0.00}}$ & $\underset{(3)}{0.41}$         & \textbf{8}                \\ \hline
Re-EC1  & 95.37                        & 3.28                        & 48.90                        & 3.50                        & 16.3                       & 10.0                       & 11.0                       & 7.3                        & $\underset{(3)}{0.55}$          & $\underset{(3)}{0.49}$          & $\underset{(1)}{0.37}$          & $\underset{(3)}{0.66}$         & {\color{blue}10} \\
2.5     & 94.87                        & 3.28                        & 49.21                        & 3.51                        & 13.0                       & 9.7                        & 12.3                       & 8.7                        & $\underset{(2)}{0.60}$          & $\underset{(3)}{0.48}$          & $\underset{(1)}{0.29}$           & $\underset{(3)}{0.66}$         & 9                         \\
5       & 95.09                        & 3.28                        & 49.58                        & 3.53                        & 14.7                       & 12.8                       & 13.3                       & 10.7                       & $\underset{(2)}{0.61}$          & $\underset{(3)}{0.45}$          & $\underset{(1)}{0.24}$          & $\underset{(3)}{0.59}$         & 9                         \\
10      & 94.37                        & 3.27                        & 49.07                        & 3.52                        & 9.3                        & 9.3                        & 13.0                       & 10.3                       & $\underset{(3)}{0.66}$          & $\underset{(3)}{0.49}$          & $\underset{(1)}{0.23}$          & $\underset{(3)}{0.64}$         & {\color{blue}10} \\
15      & 94.94                        & 3.28                        & 49.56                        & 3.53                        & 11.0                       & 10.8                       & 12.7                       & 11.0                       & $\underset{(2)}{0.63}$          & $\underset{(3)}{0.43}$          & $\underset{(1)}{0.25}$           & $\underset{(3)}{0.63}$         & 9                         \\
20      & 95.23                        & 3.29                        & 50.04                        & 3.53                        & 13.7                       & 12.8                       & 16.0                       & 12.3                       & $\underset{(2)}{0.63}$          & $\underset{(3)}{0.47}$          & $\underset{(1)}{0.21}$          & $\underset{(3)}{0.61}$         & 9                         \\ \hline
\end{tabular}
}
\end{table}
\end{center}


\vspace{0.5cm}
{\centering\section{\normalsize CONCLUSION} \label{conclusion_section}}
\noindent
We have presented a novel method, called QFHS, for generating multi-step ahead forecasts of VaR and ES. The proposed method can be seen as a generalization of the well-known filtered historical simulation, where the Gaussian QL function is replaced by the quantile loss and resampling is based on residuals scaled by the series of conditional quantiles rather than variances. The QFHS extends the application of simple CAViaR models to the generation of ES and multi-step ahead forecasts in a simple and flexible manner. 

Furthermore, the QFHS brings additional flexibility to the standard CAViaR modelling framework by allowing $\alpha_{\text{est}}$, the quantile level used to estimate the CAViaR model, to be decoupled from the target risk level $\alpha_{0}$. Our empirical findings suggest that the optimal $\alpha_\text{est}$ is always well above the target risk level $[0.01, 0.025]$. More specifically, the numerical experiments presented in the paper consistently show that when $\alpha_{\text{est}}$ is within $[0.05, 0.15]$, the simulated and empirical performance of the QFHS predictors is generally optimal or close to optimal under different criteria (i.e. risk prediction accuracy, estimation SEs). 

Finally, in risk forecasting applications to both real and simulated data, we find that QFHS-based predictors generated by choosing $\alpha_{\text{est}}$ in $[0.05, 0.15]$ tend to outperform their FHS-based counterparts for high kurtosis and skewed data, as in the case of individual stocks. Therefore, following the above discussion and based on our empirical results, as a general recommendation for practitioners interested in applying QFHS to real stock market data, we would suggest choosing $\alpha_{\text{est}}$ in the range $[0.05, 0.15]$, for typical choices of $\alpha_0\leq 0.025$, such as those considered in the paper.


\noindent
	\bibliographystyle{chicago}
	\bibliography{main}

\section*{Disclosure of Funding}
Giuseppe Storti and Antonio Naimoli acknowledge financial support under the National Recovery and Resilience Plan (NRRP), Mission 4, Component 2, Investment 1.1, Call for tender No. 104 published on 2.2.2022 by the Italian Ministry of University and Research (MUR), funded by the European Union – NextGenerationEU– Project Title: Methodological and computational issues in large-scale time series models for economics and finance (ID: 20223725WE) – CUP  D53D 2300610 0006 - Grant Assignment Decree No. n. 967 of 30/06/2023.


\clearpage








\begin{appendices}
\renewcommand{\thesection}{\normalsize\Alph{section}}

\vspace{0.5cm}
{\centering \section*{\normalsize APPENDIX} \label{sec:apps} }
\noindent

{\centering \section*{\normalsize  ADDITIONAL SIMULATION RESULTS} \label{sec:appA} }

\setcounter{figure}{0} \renewcommand{\thefigure}{A.\arabic{figure}}
\setcounter{table}{0} \renewcommand{\thetable}{A.\arabic{table}}
\setcounter{subsection}{0} \renewcommand{\thesubsection}{A.\arabic{subsection}}

\begin{figure}[h!]
		\centering
		\caption{Normalized Monte Carlo standard errors (raw data) for $\beta$ (top) and $\gamma$ (bottom) with respect to different levels of $\alpha$. The patterns are generated for different degrees of freedom $\nu$ and setting the skewness to $\xi=-0.05$. \\}
		\includegraphics[width=\textwidth]{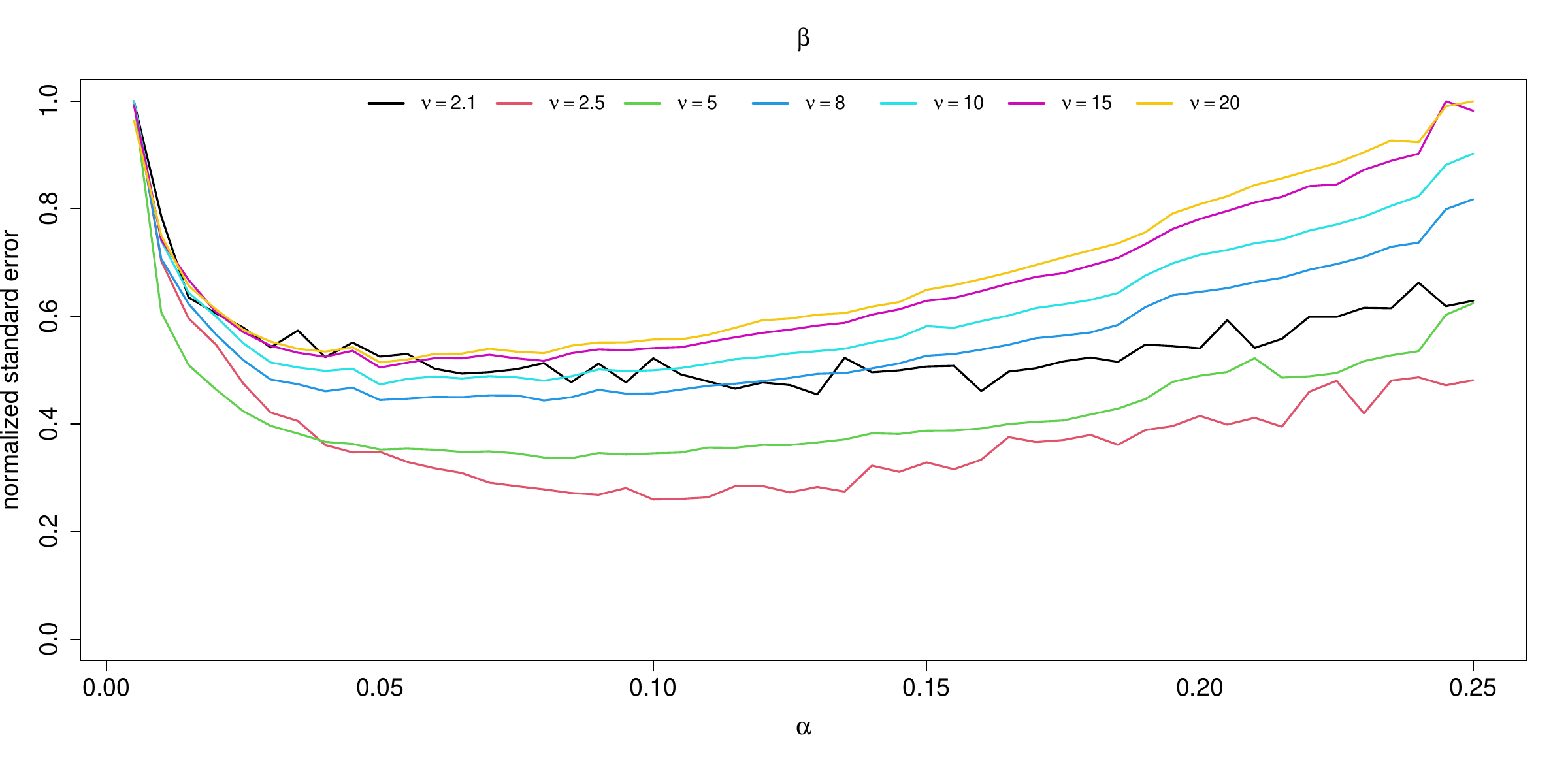}\\
        		\includegraphics[width=\textwidth]{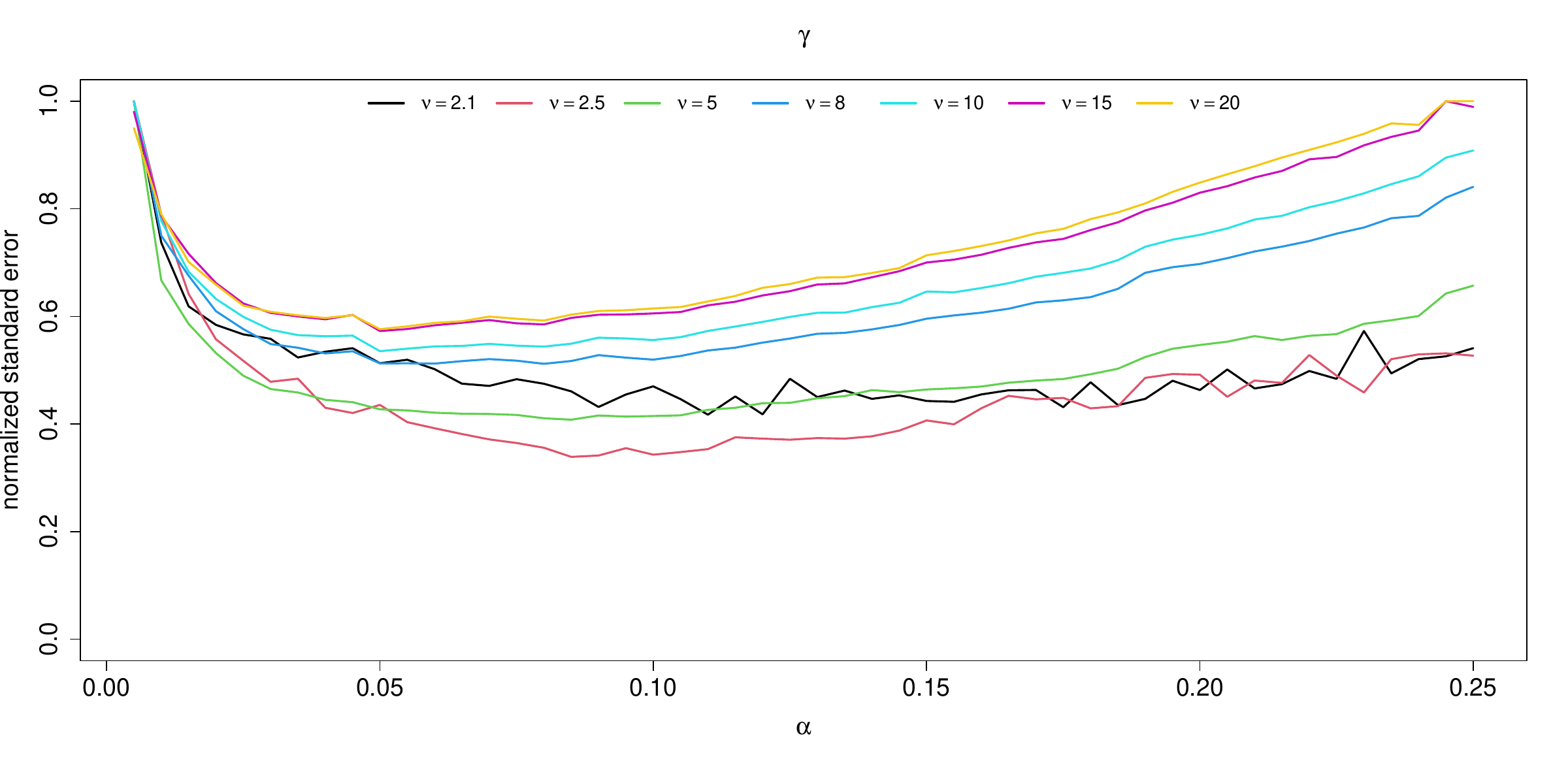}
                		\label{fig:all_beta_gamma_raw}

	\end{figure}

\begin{figure}[h!]
\centering
\caption{Normalized Monte Carlo standard errors (loess smoothed data) for $\beta$ (top) and $\gamma$ (bottom) with respect to different levels of $\alpha$. The patterns are generated for different degrees of freedom $\nu$ and setting the skewness to $\xi=-0.05$. \\}
		\includegraphics[width=\textwidth]{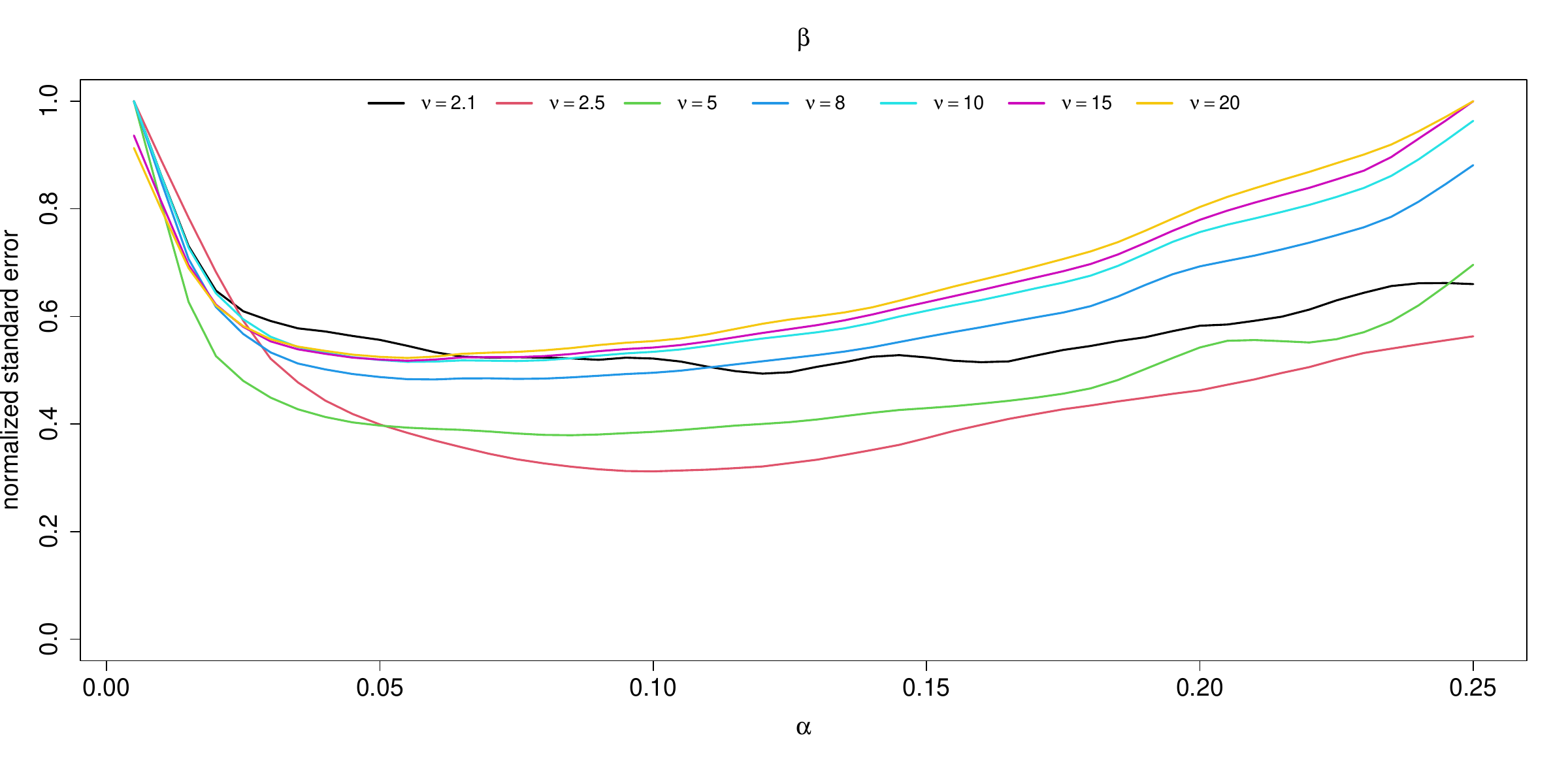}\\
       		\includegraphics[width=\textwidth]{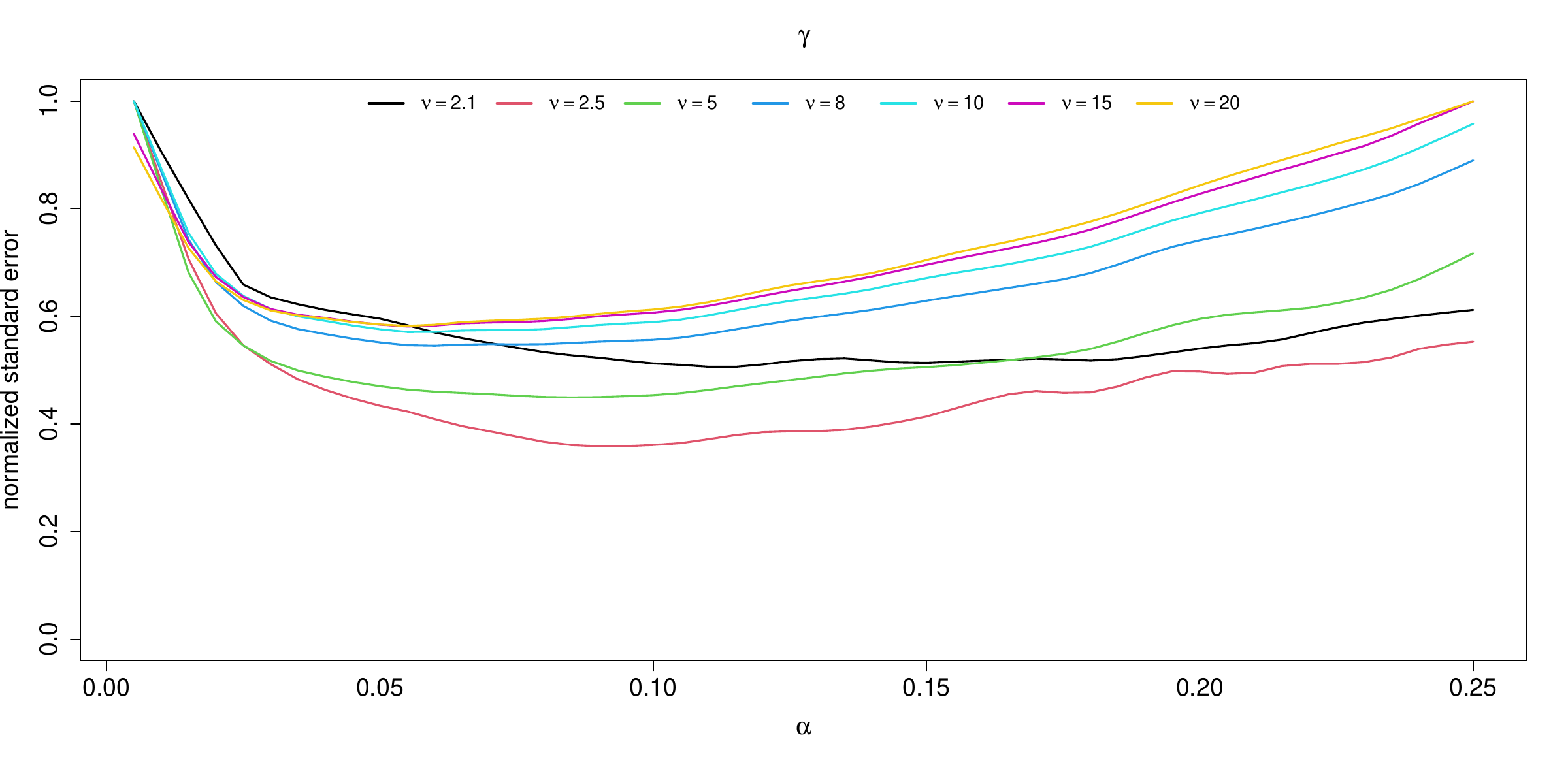}
            \label{fig:all_beta_gamma_loess}

	\end{figure}

In this section, we illustrate the results of the simulation study introduced in the second part of Section \ref{ss:secaviar}. First, in Figure \ref{fig:all_beta_gamma_raw} the Monte Carlo standard errors for $\beta$ (top) and $\gamma$ (bottom) are plotted against the value of the estimation $\alpha$, for different values of $\nu$ and fixed $\xi=-0.05$. Despite the obvious effects of finite sample estimation noise, the patterns are very similar to those shown in Figure \ref{fig:se_norm_omega}, indicating that extreme (too low) values of $\alpha$ should be avoided. As expected, the same applies to high values of $\alpha$, far from the tail. To mitigate measurement uncertainties, we also report results for the $\alpha$ level minimizing standard errors obtained from the  
fitting of a local polynomial regression (LOESS) with a smoothing parameter selected by a bias-corrected Akaike Information Criterion.The loess smoothed version of the plots shown in Figure \ref{fig:all_beta_gamma_loess},
makes the similarity with the pattern in Figure \ref{fig:se_norm_omega} even more obvious.

\begin{figure}
		\centering
		\caption{Monte Carlo standard errors for $\beta$ and $\gamma$ with respect to different levels of $\alpha$. The patterns are generated by setting the degrees of freedom $\nu\in (2.1,2.5,5)$ and the skewness $\xi=-0.05$. Black line: raw data. Red line: loess-smoothed data. \\}
		\includegraphics[width=0.9\textwidth]{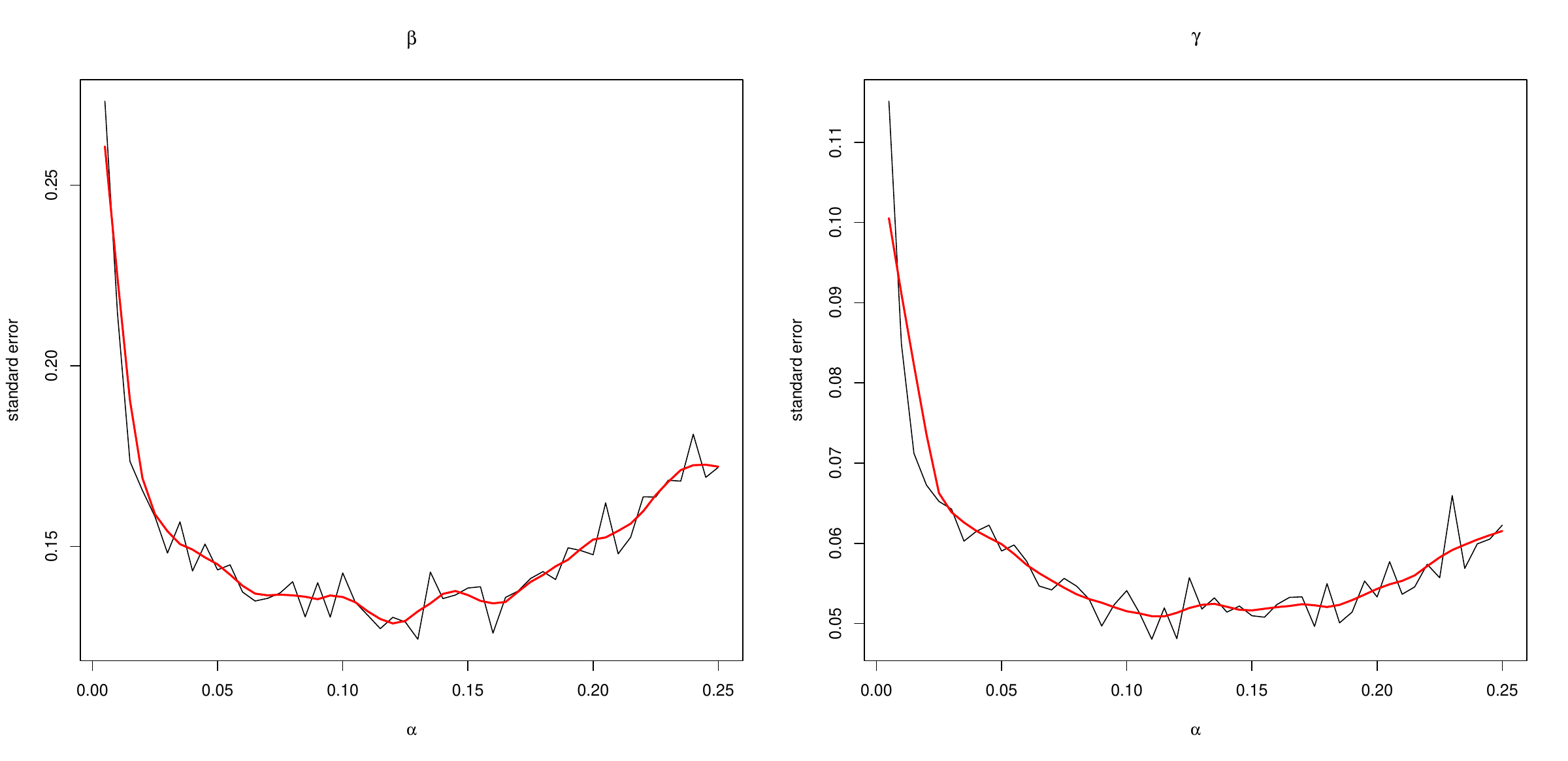}\\
        		\includegraphics[width=0.9\textwidth]{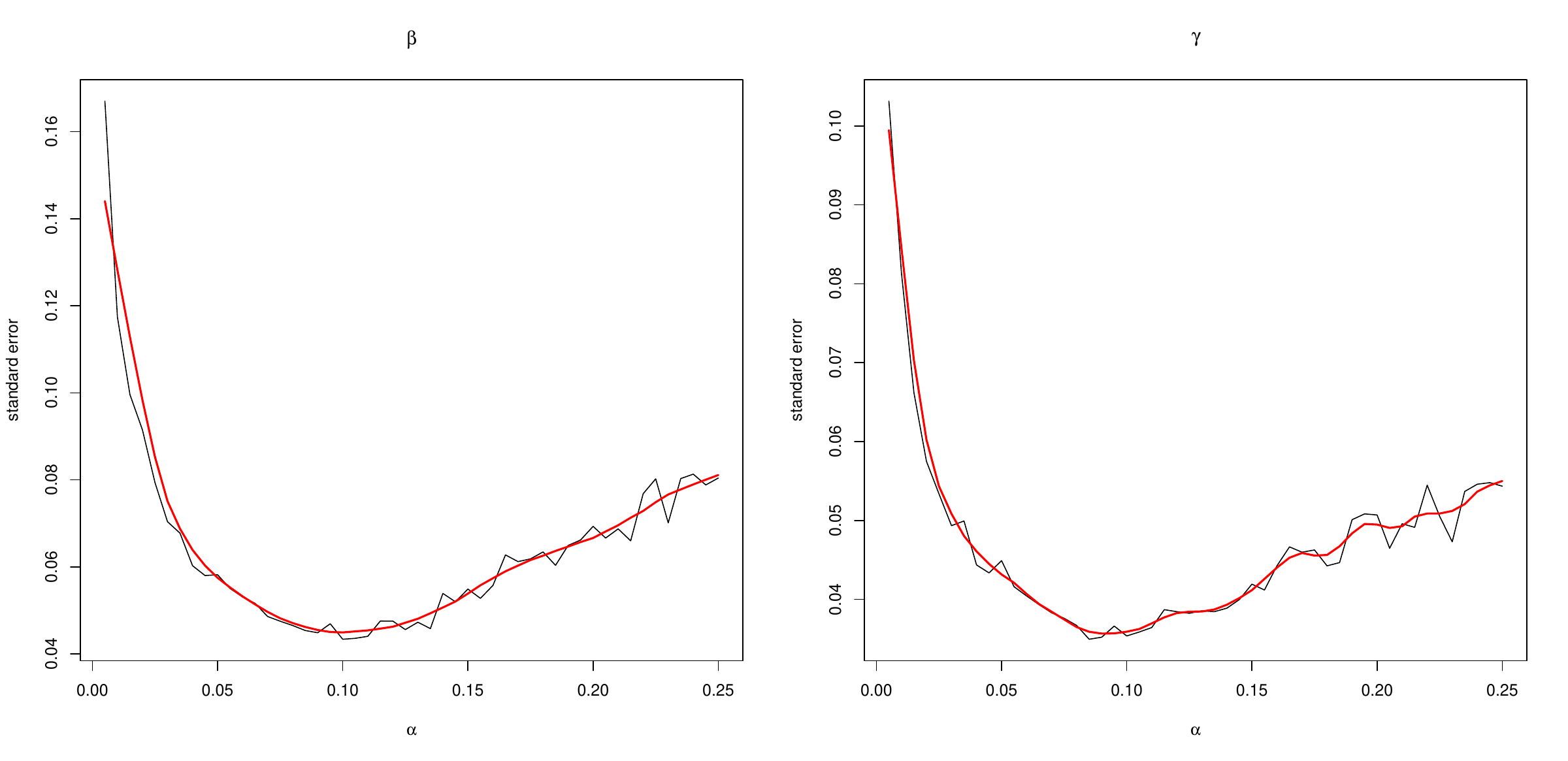}\\
		\includegraphics[width=0.9\textwidth]{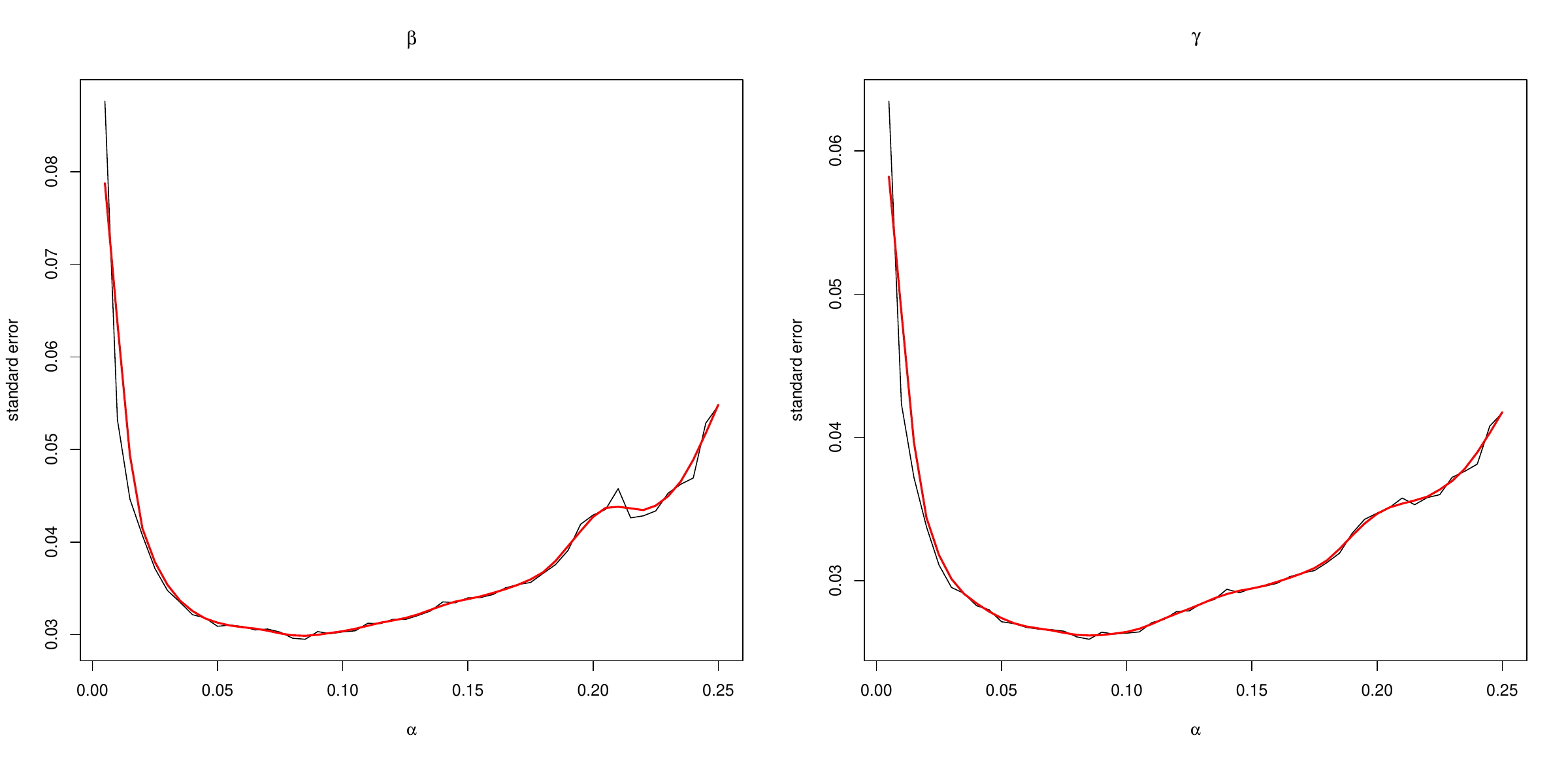}
		\label{fig:se_df_xi-0.05}

	\end{figure}

Furthermore, for the convenience of the reader, a selection of patterns, for $\nu\in(2.1,2.5,5)$ and $\xi=-0.05$, are shown in Figure \ref{fig:se_df_xi-0.05}, where the raw patterns are compared with their smoothed counterparts. In all cases the optimal value of $\alpha$ is close to 0.10. It should also be noted that while the steepness of the curve tends to increase as we approach the extremes of the range of $\alpha$ values, it is moderate near the minimum, meaning that values of $\alpha$ that are reasonably close to the minimum do not lead to significant losses in efficiency.


Table \ref{tab:mean_beta_gamma_sim} reports the average level of $\alpha$ that minimizes standard errors (raw data) for the parameters $\beta$ and $\gamma$, under different combinations of the degrees of freedom $\nu$ and skewness $\xi$. It should be noted that, unlike the simulation study based on the theoretical asymptotic standard errors, in this setting the optimal $\alpha$ could be different for $\beta$ and $\gamma$. Hence, Table \ref{tab:mean_beta_gamma_sim} reports the average of optimal estimation $\alpha$ values obtained for $\gamma$ and $\beta$, respectively. As a robustness check, the optimal values of $\alpha$ computed from the loess smoothed standard errors are reported in Figure \ref{tab:mean_beta_gamma_sim_loess}

\begin{table}[htb]
\centering
	\caption{Average (across $\gamma$ and $\beta$) optimal $\alpha$ levels minimizing Monte Carlo standard errors (raw data), under different combinations of $\nu$ and  $\xi$. \\}
\label{tab:mean_beta_gamma_sim}
\begin{tabular}{c|ccccccc|}
\cline{2-8}                              & \multicolumn{7}{c|}{$\xi$}                                   \\ \hline
\multicolumn{1}{|c|}{$\nu$}    & -0.15  & -0.10  & -0.05  & 0      & 0.05   & 0.10   & 0.15   \\ \hline
\multicolumn{1}{|c|}{2.1}      & 0.1025 & 0.1200 & 0.1200 & 0.1275 & 0.1725 & 0.1450 & 0.1675 \\
\multicolumn{1}{|c|}{2.5}      & 0.1000 & 0.1025 & 0.0925 & 0.0950 & 0.0975 & 0.1050 & 0.1175 \\
\multicolumn{1}{|c|}{5}        & 0.0750 & 0.0800 & 0.0850 & 0.0850 & 0.0900 & 0.0825 & 0.0900 \\
\multicolumn{1}{|c|}{8}        & 0.0700 & 0.0800 & 0.0800 & 0.0675 & 0.0800 & 0.0675 & 0.0650 \\
\multicolumn{1}{|c|}{10}       & 0.0725 & 0.0500 & 0.0500 & 0.0500 & 0.0550 & 0.0525 & 0.0550 \\
\multicolumn{1}{|c|}{15}       & 0.0500 & 0.0500 & 0.0500 & 0.0500 & 0.0550 & 0.0525 & 0.0525 \\
\multicolumn{1}{|c|}{20}       & 0.0500 & 0.0500 & 0.0500 & 0.0500 & 0.0525 & 0.0500 & 0.0525 \\
\multicolumn{1}{|c|}{$\infty$} & -      & -      & -      & 0.0500 & -      & -      & -      \\ \hline
\end{tabular}%
\end{table}

The loess smoothed results (Table \ref{tab:mean_beta_gamma_sim_loess}) are reasonably close to the un-smoothed results, except for the extreme case where $\nu=2.1$. In general, both tables show a pattern similar to that observed in Table \ref{tab:param_alpha_min} with a tendency to return optimal $\alpha$ levels slightly lower than those based on asymptotic standard errors.

	\begin{table}[htb]
		\centering
		\caption{Average (across $\gamma$ and $\beta$) optimal $\alpha$ levels minimizing Monte Carlo standard errors (loess smoothed), under different combinations of $\nu$ and  $\xi$. \\}
		\label{tab:mean_beta_gamma_sim_loess}
			\begin{tabular}{c|ccccccc|}
				\cline{2-8}
				& \multicolumn{7}{c|}{$\xi$}                                   \\ \hline
				\multicolumn{1}{|c|}{$\nu$}    & -0.15  & -0.10  & -0.05  & 0      & 0.05   & 0.10   & 0.15   \\ \hline
				\multicolumn{1}{|c|}{2.1}      & 0.1150 & 0.1150 & 0.1175 & 0.1425 & 0.1650 & 0.1650 & 0.1750 \\
				\multicolumn{1}{|c|}{2.5}      & 0.0875 & 0.0975 & 0.0950 & 0.0950 & 0.0950 & 0.1000 & 0.1225 \\
				\multicolumn{1}{|c|}{5}        & 0.0775 & 0.0850 & 0.0850 & 0.0850 & 0.0900 & 0.0950 & 0.1000 \\
				\multicolumn{1}{|c|}{8}        & 0.0675 & 0.0675 & 0.0600 & 0.0600 & 0.0675 & 0.0700 & 0.0675 \\
				\multicolumn{1}{|c|}{10}       & 0.0600 & 0.0600 & 0.0550 & 0.0575 & 0.0600 & 0.0675 & 0.0650 \\
				\multicolumn{1}{|c|}{15}       & 0.0550 & 0.0550 & 0.0550 & 0.0550 & 0.0550 & 0.0550 & 0.0550 \\
				\multicolumn{1}{|c|}{20}       & 0.0550 & 0.0550 & 0.0550 & 0.0550 & 0.0550 & 0.0550 & 0.0550 \\
				\multicolumn{1}{|c|}{$\infty$} & -      & -      & -      & 0.0500 & -      & -      & -      \\ \hline
			\end{tabular}%
	\end{table}

\begin{figure}[h!]
		\centering
		\caption{Evolution of the optimal $\alpha$ computed as the average of the minimum $\alpha$ for $\gamma$ and $\beta$ (raw data) over the skewness coefficient $\xi$, for different values of $nu$  (top), and the degrees of freedom $\nu$, for different values of $\xi$ (bottom).}
		\includegraphics[width=\textwidth]
        {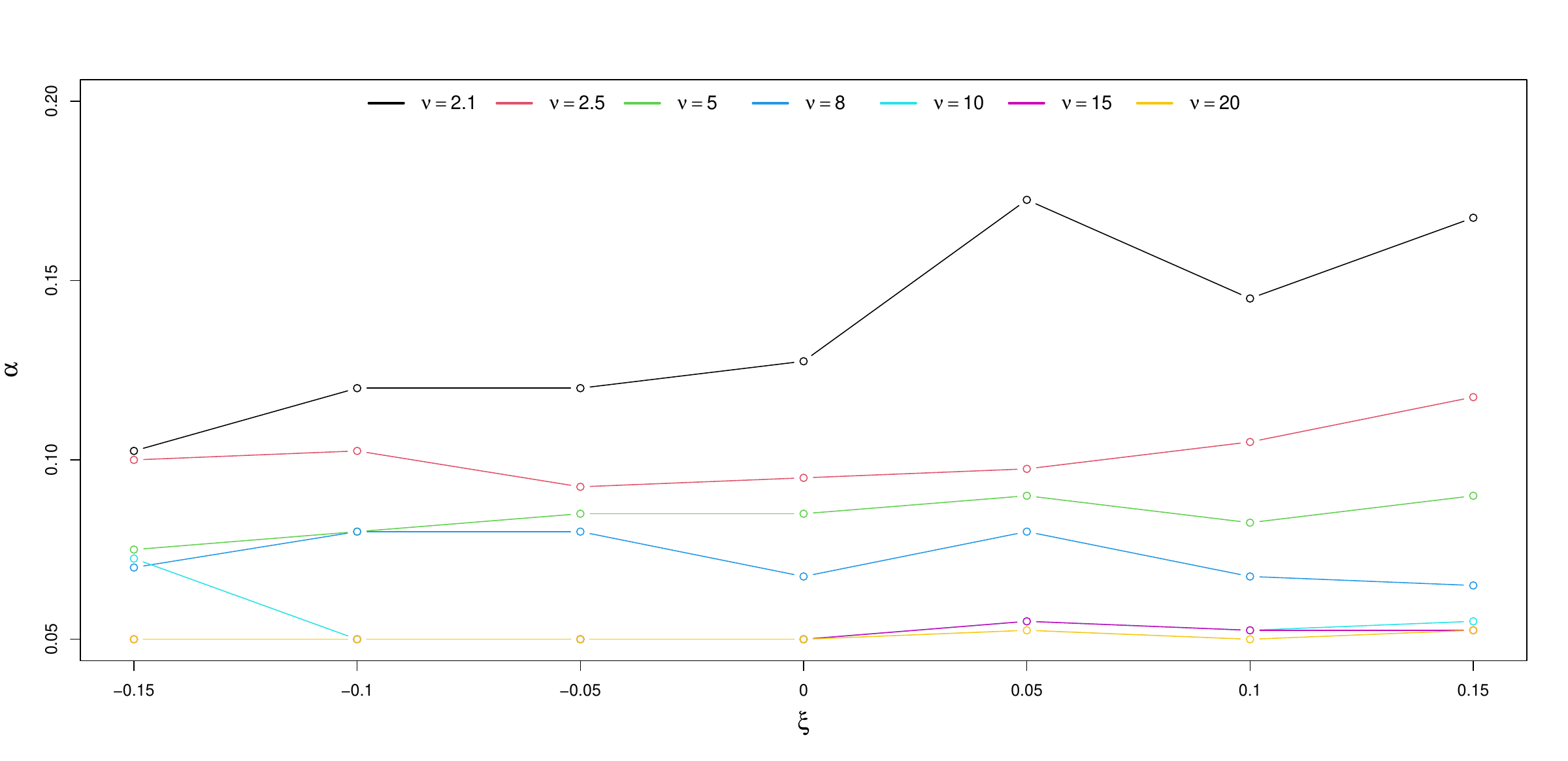}\\		\includegraphics[width=\textwidth]
        {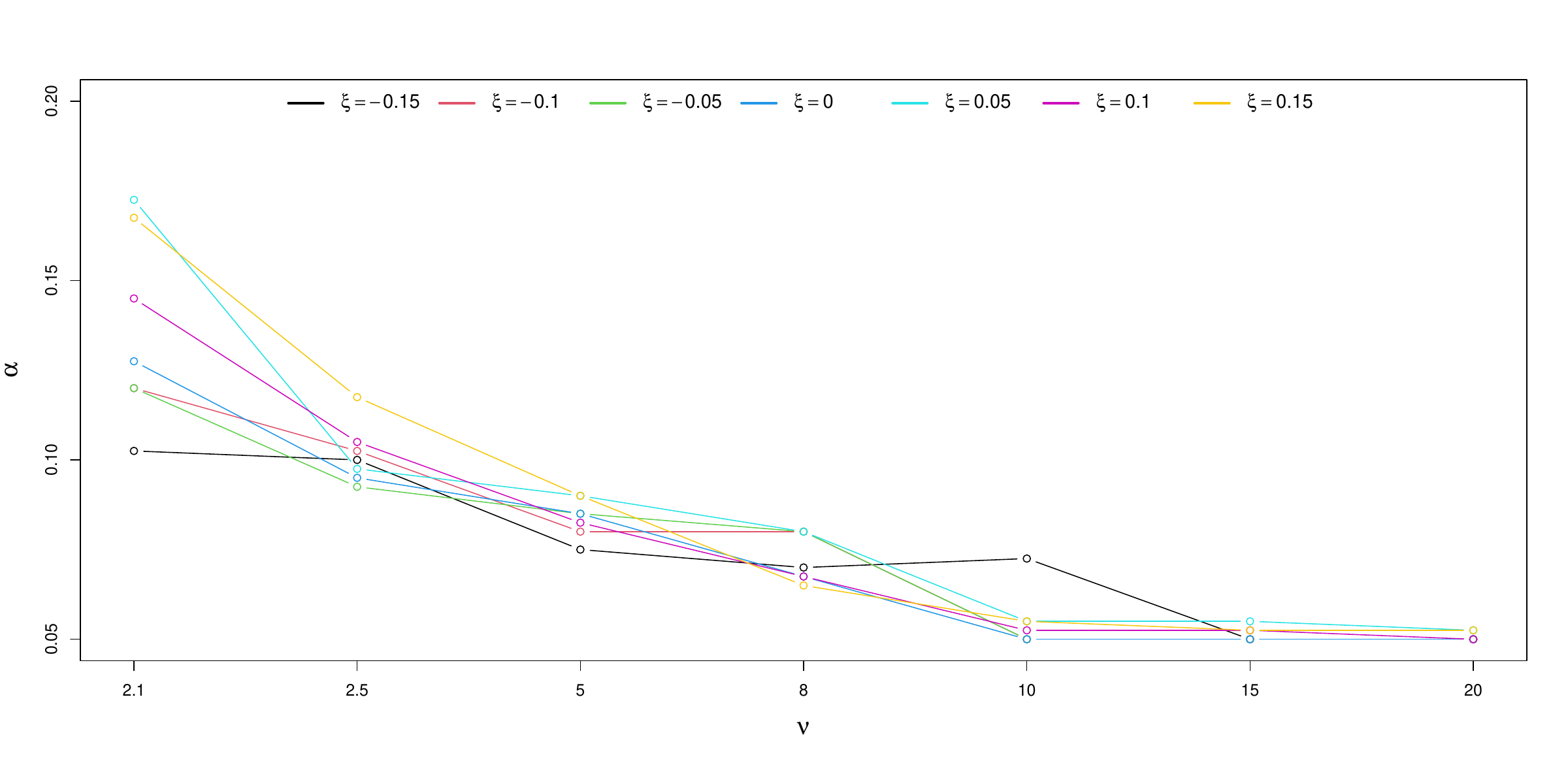}
		\label{fig:alpha_fixed_xi_raw}

	\end{figure}

Finally, the plots in Figure \ref{fig:alpha_fixed_xi_raw} allow for a more effective visualization of the pattern of the dependence of the optimal $\alpha$ on $\nu$ (for fixed $\xi$) and $\xi$ (for fixed $\nu$), respectively. Again, taking into account the effect of estimation noise, these patterns are quite similar to those observed in Figure \ref{fig:plot2}.

\end{appendices}

\end{document}